\documentclass[twocolumn, tighten, twocolappendix]{aastex63}

\usepackage{amsmath}
\usepackage{textcomp}
\usepackage{placeins}

\accepted{June 12, 2020} 
\submitjournal{\apj}

\graphicspath{{./}{figures/}}

\begin{document}

\title{LOW ALBEDO SURFACES OF LAVA WORLDS}

\correspondingauthor{Zahra Essack}
\email{zessack@mit.edu}

\author[0000-0002-2482-0180]{Zahra Essack}
\affiliation{Department of Earth, Atmospheric and Planetary Sciences, Massachusetts Institute of Technology, Cambridge, MA 02139, USA}
\affiliation{Kavli Institute for Astrophysics and Space Research, Massachusetts Institute of Technology, Cambridge, MA 02139, USA}

\author[0000-0002-6892-6948]{Sara Seager}
\affiliation{Department of Earth, Atmospheric and Planetary Sciences, Massachusetts Institute of Technology, Cambridge, MA 02139, USA}
\affiliation{Department of Physics, and Kavli Institute for Astrophysics and Space Research, Massachusetts Institute of Technology, Cambridge, MA 02139, USA}
\affiliation{Department of Aeronautics and Astronautics, MIT, 77 Massachusetts Avenue, Cambridge, MA 02139, USA}

\author[0000-0003-3500-2770]{Mihkel Pajusalu}
\affiliation{Department of Earth, Atmospheric and Planetary Sciences, Massachusetts Institute of Technology, Cambridge, MA 02139, USA}
\affiliation{Tartu Observatory, University of Tartu, Observatooriumi 1, 61602 Toravere, Estonia}



\begin{abstract}

Hot super Earths are exoplanets with short orbital periods ($<$ 10 days), heated by their host stars to temperatures high enough for their rocky surfaces to become molten. A few hot super Earths exhibit high geometric albedos ($>$ 0.4) in the \textit{Kepler} band (420-900 nm). We are motivated to determine whether reflection from molten lava and quenched glasses (a product of rapidly cooled lava) on the surfaces of hot super Earths contributes to the observationally inferred high geometric albedos. We experimentally measure reflection from rough and smooth textured quenched glasses of both basalt and feldspar melts. For lava reflectance values, we use specular reflectance values of molten silicates from  non-crystalline solids literature. Integrating the empirical glass reflectance function and non-crystalline solids reflectance values over the dayside surface of the exoplanet at secondary eclipse yields an upper limit for the albedo of a lava-quenched glass planet surface of $\sim $0.1. We conclude that lava planets with solid (quenched glass) or liquid (lava) surfaces have low albedos. The high albedos of some hot super Earths are most likely explained by atmospheres with reflective clouds (or, for a narrow range of parameter space, possibly Ca/Al oxide melt surfaces). Lava planet candidates in \textit{TESS} data can be identified for follow-up observations and future characterization.

\end{abstract}




\section{Introduction} \label{sec:intro}

With over 4000 exoplanets discovered to date\footnote{https://exoplanetarchive.ipac.caltech.edu/docs/counts\_detail.html}, exoplanet research is moving towards characterizing classes of exoplanets. The pioneering \textit{Kepler} telescope transformed our understanding of exoplanets with discoveries of new planet categories and planetary systems. Some of \textit{Kepler's} discoveries include: terrestrial-sized planets are common; mini Neptunes/super Earths are the most common planet size; and compact multiple planet systems. In this work we pursue hot super Earths.

Super Earths are a subset of the exoplanet population with masses of 1$-$10~M$_{\oplus }$. The term super Earth used in this context refers to a primarily rocky planet without a significant envelope \citep{seager2007mass}. Hot super Earths are exoplanets with short orbital periods ($<$ 10 days) that are strongly irradiated by their host stars, which leads to high surface temperatures. Because temperatures on some hot super Earths are high enough for their rocky surfaces to become molten, hot super Earths can be further categorized into lava-ocean planets \citep{leger2009transiting, leger2011extreme}.

Lava-ocean planets are compelling to study for the insight they provide into the behaviors of materials at extreme temperatures, volatile cycling, and Earth's early history.

\subsection{Lava-Ocean Planets} \label{subsec:lavaoceanplanet}

Lava-ocean planets are short period ($<$ 10 days), rocky planets (R$_{p}$ $<$ 1.6~R$_{\oplus }$)\citep{rogers2015most} that are expected to be tidally locked, and have their orbits circularized by tidal interactions due to their close proximity to their host stars.

Lava-ocean planets have been previously theorized. The first two potential lava-ocean planets were Kepler-10 b (R$_{p}$ = 1.47$_{-0.02}^{+0.03}$~R$_{\oplus }$, P = 0.8374907 $\pm $ $2\times 10^{-7}$~days) \citep{dumusque2014kepler} and CoRoT-7 b (R$_{p}$ = 1.58 $\pm $0.1~R$_{\oplus }$, P = 0.85359 $\pm \ 3\times 10^{-5}$~days) \citep{leger2011extreme}. 

\begin{figure}[htb!]
\vspace*{5mm}
\includegraphics[width=\columnwidth]{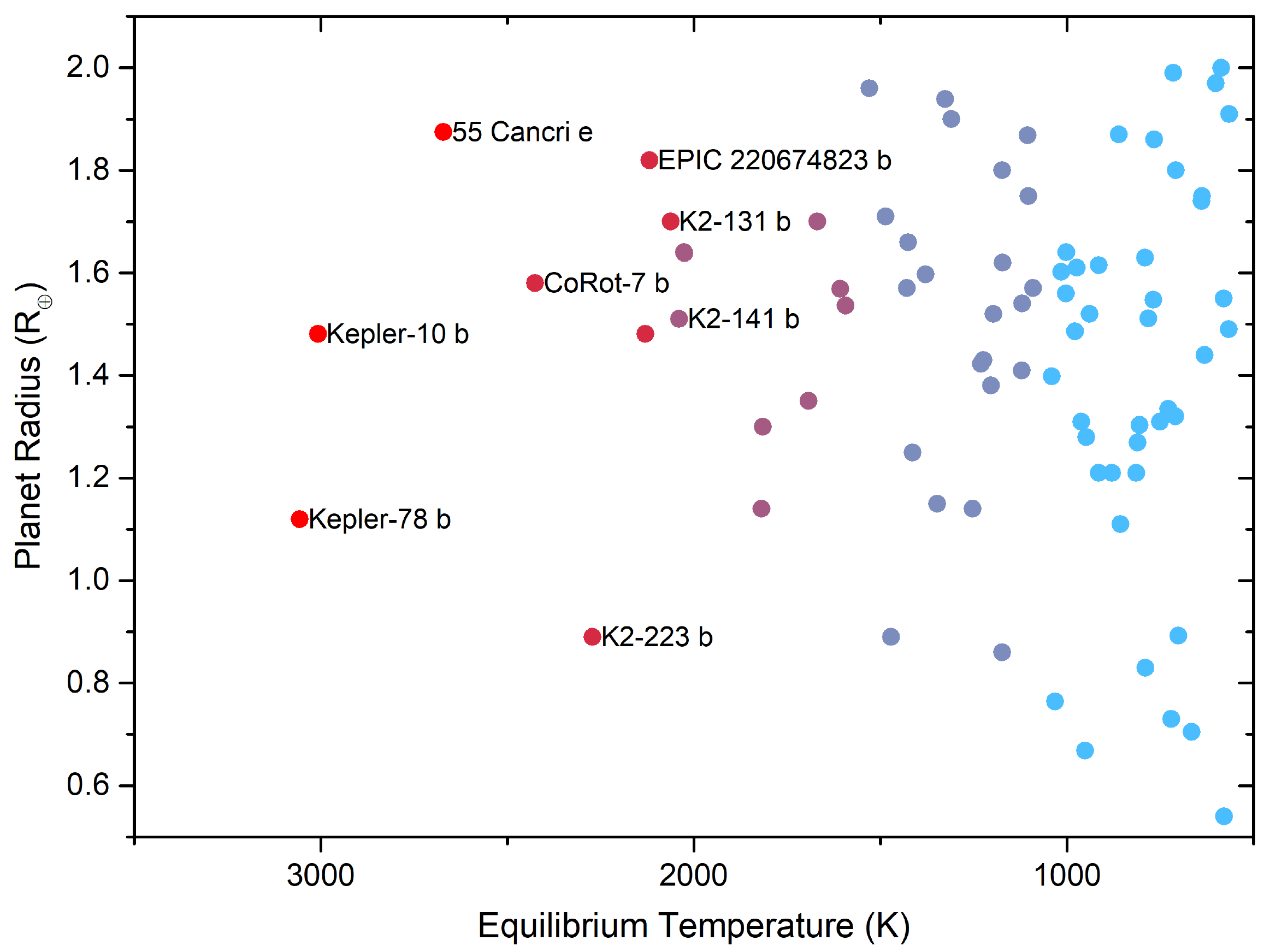}
\caption{Planet radius vs. equilibrium temperature of lava-ocean planet candidates. Planet color corresponds to equilibrium temperature. Lava-ocean planets are short-period ($<$ 10 days), rocky planets, with surface temperatures high enough to melt the crust on the dayside hemisphere ($>$ 850 K). Planets with R$_{p}$ $>$ 1.6~R$_{\oplus }$ are likely to have volatile envelopes (\citet{rogers2015most, fulton2017california}), so planets must have R$_{p}$ $<$ 1.6~R$_{\oplus }$ to conform with the lava-ocean planet definition. \label{fig:RvsTemp}}
\end{figure}

Lava-ocean planets are expected to have cloudless, low pressure atmospheres consisting of rocky vapors, large day-night temperature differences, and an ocean of molten refractory rocks on the strongly irradiated dayside of the planet \citep{leger2011extreme}. Surface temperatures on lava-ocean planets must be greater than $\sim $850 K in order to sustain the molten lava ocean on the dayside hemisphere, assuming a planetary crust composition similar to Earth\footnote{Assuming an Earth-like crustal composition, the lower limit of 850 K on the surface temperature of lava-ocean exoplanets is related to the melting points of silicate rocks on Earth. Silicate rocks begin melting at 850 K, and all silicate rocks are molten above 1473 K \citep{lutgens2014essentials}.}. Currently known lava-ocean planet candidates are shown in Figure \ref{fig:RvsTemp}.

\subsection{Secondary Eclipse Measurements} \label{subsec:transitseclipses}

Secondary eclipse measurements are the best way to identify and characterize lava-ocean planets in the near future. Due to their close proximity to their host stars, short-period exoplanets receive large amounts of incident stellar flux which increases their detectable reflected and thermal emission, making them particularly well-suited to secondary eclipse measurements. Short orbital periods also enable several eclipses to be captured within a short observing time frame \citep{alonso2018characterization}.  

Indeed, a few potential lava-ocean planets show a significant secondary eclipse signal. For some planets, the signal is interpreted as a high geometric albedo in the \textit{Kepler} band (420$-$900 nm). The hot super Earths considered have median geometric albedos ranging from 0.16 to 0.30 \citep{demory2014albedos}. Further, there are a small number of exoplanets including Kepler-10 b and Kepler-21 b that have geometric albedos $>$~0.4 \citep{batalha2011kepler, demory2014albedos}.

A degeneracy exists between reflected light and thermal emission when interpreting broadband observations of the secondary eclipse depth $\left( \frac{F_{p}}{F_{s}}\right)$ of short period exoplanets. The total planetary flux, $F_{p}$, is a combination of thermal emission and reflected light. The reflected light component of the secondary eclipse depth is given by:

\begin{equation}
\frac{F_{p}}{F_{s}}=A_{g}\left( \frac{R_{p}}{a}\right) ^{2}\phi (\alpha )%
\text{ ,}  \label{eqn1}
\end{equation}

\noindent where $F_{s}$ is the stellar flux, $A_{g}$ is the geometric albedo, $R_{p}$ is the planet radius, $a$ is the semi-major axis/orbital distance, and $\phi (\alpha )$ is the phase function (Section \ref{modelmethod}).

\begin{figure*}[htb!]
\plotone{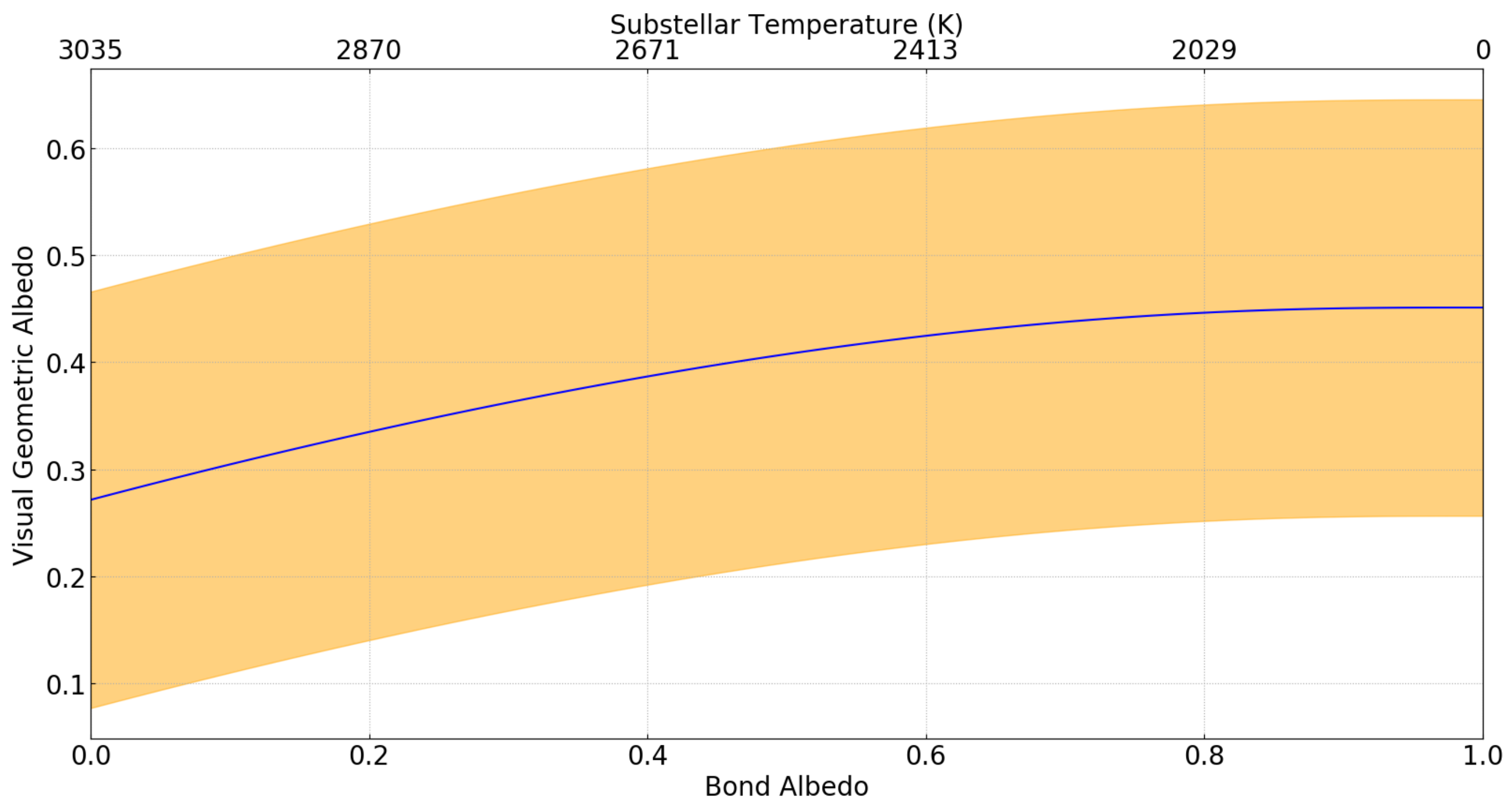}
\caption{Range of substellar temperatures, geometric albedo values, and bond albedo values for Kepler-10 b based on its measured secondary eclipse depth of 5.8 $\pm$ 2.5 ppm. The shaded region is the 68\% confidence region for the geometric albedo (adapted from \citet{malavolta2018ultra}). This illustrates the degeneracy between albedo and temperature for a measured secondary eclipse depth. Kepler-10 b has a large range of substellar temperatures (850 - 3035 K) high enough to sustain molten lava on the surface, covering most of the possible values on the plot.\bigskip 
\label{fig:AlbedoDegeneracy}}
\end{figure*}

Lava-ocean exoplanets and other short period planets are typically heated to temperatures of 1000$-$3000~K, which causes an observable thermal flux at visible wavelengths, as well as a reflected light component. This leads to a range of geometric albedos and planet substellar temperatures that could result in the measured secondary eclipse depth (Figure \ref{fig:AlbedoDegeneracy}). 

The total planetary flux needs to be decontaminated from thermal emission before calculating the geometric albedo by removing the estimated blackbody thermal emission contribution \citep{demory2014albedos}. The thermal emission can be estimated by the blackbody thermal emission contribution:

\begin{equation}
\frac{F_{p}}{F_{s}}=\left( \frac{R_{p}}{R_{s}}\right) ^{2}\frac{B(\lambda
,T_{p})}{B(\lambda ,T_{s})}\text{ ,}  \label{eqn2}
\end{equation}

\noindent where $R_{s}$ is the radius of the star, and $B(\lambda ,T_{s})$ and $B(\lambda ,T_{p})$ are the blackbody emissions (Planck distribution functions) of the star and the planetary dayside at brightness temperatures of $T_{s}$ and $T_{p}$, respectively \citep{alonso2018characterization}.

\subsection{Quenched Planetary Glasses as a Possible Source of Reflection} \label{quenchedglass}

In this study, we aim to determine whether reflection from molten lava and quenched glasses (a product of rapidly cooled lava) on the surfaces of hot super Earths may be sources of reflected light that contribute to the high geometric albedos.

Quenched glasses, more specifically silicate glasses, are amorphous phases naturally formed through volcanism and hypervelocity impacts. A study by \citet{cannon2017spectral} measured VNIR spectra of synthetic glasses with different compositions. A small subset of the synthetic glasses formed under extremely low oxygen conditions and containing little/no iron had reflectance values as high as 75\% at certain visible wavelengths. If cooled rapidly in relation to wind action, quenched glasses can have a very smooth texture, motivating us to consider reflection from their surfaces.

Lava-ocean exoplanets are assumed to have low pressure atmospheres with little to no volatile species (e.g. H, C, N) due to their close proximity to their host stars, which results in effective atmospheric stripping. Due to the low pressure atmosphere, we assume a steep temperature gradient on these planets, i.e., there is a rapid decrease in temperature as distance increases outward from the substellar point. 

We assume that the planetary crust is melted to form lava at the substellar point. As the lava flows away from the substellar point, it rapidly cools to form quenched glass. We aim to measure the reflectivity of quenched glasses to determine if it significantly contributes to the high geometric albedos of some hot super Earths. 

In Section \ref{sec:methods}, we describe our quenched glass sample preparation, experimental method to measure reflection from quenched glasses, and model the geometric albedo on the dayside hemisphere of the exoplanet. In Section \ref{sec:results}, we present our experimental data, model fit results, and calculated geometric albedo values for reflection from lava and quenched glasses on the planetary surface. In Section \ref{sec:discussion}, we discuss several topics including: reflectance simulations to support our experimental albedo estimates; the implications of our results for lava and quenched glasses as sources of reflection on lava-ocean exoplanets; how quenched glasses formed under lava-ocean exoplanet conditions may differ from the quenched glasses in this study; a description of other potential high albedo surfaces; and  reflection in the atmospheres of hot super Earths. We discuss the challenges involved in measuring reflection from molten lava in the laboratory, and conclude with a description of the possibilities of lava-ocean planet candidates in \textit{TESS} data for future characterization.

\section{Methods} \label{sec:methods}

We describe our sample preparation procedure and experimental setup, and outline our model calculations.

\subsection{Sample Preparation} \label{sampleprep}

Our samples of quenched glass were obtained by melting crushed basalt rock and soda feldspar powder, and allowing the melts to rapidly cool in air.

The basalt rock starting material used is ancient basalt ($\sim $1.2 billion years old) from lava flows that were erupted in the Mid-Continent Rift, sourced from the Chengwatana Formation in Dresser Trap Rock Quarry, Wisconsin \citep{wirth1997chengwatana}. The feldspar starting material used is 200 mesh soda feldspar from Spruce Pine, North Carolina (Table \ref{TableCompositions}).

\begin{center}
\begin{table}[htb!]
	
	\begin{tabular}{|c|c|c|}
		\hline
		 wt. \% & Basalt & Feldspar \\ \hline
		SiO$_{2}$ & 49.03$\pm $\ 1.77 & 68.60 $\pm $\ 6.52 \\ \hline
		Fe$_{2}$O$_{3}$ & 14.82 $\pm $\ 1.15 & 0.06 $\pm $\ 0.002 \\ \hline
		Al$_{2}$O$_{3}$ & 14.50 $\pm $\ 0.91 & 18.50 $\pm $\ 1.04 \\ \hline
		Na$_{2}$O & 2.93 $\pm $\ 0.97 & 6.50 $\pm $\ 0.60 \\ \hline
		K$_{2}$O & 1.07 $\pm $\ 0.47 & 4.10 $\pm $\ 0.25 \\ \hline
		CaO & 8.24 $\pm $\ 0.95 & 1.50 $\pm $\ 0.15 \\ \hline
		MgO & 5.52 $\pm $\ 1.14 & $-$ \\ \hline
		MnO & 0.20 $\pm $\ 0.02 & $-$ \\ \hline
		FeO & 0.00 & $-$ \\ \hline
		P$_{2}$O$_{5}$ & 0.26 $\pm $\ 0.07 & $-$ \\ \hline
		Cr$_{2}$O$_{3}$ & 0.02 $\pm $\ 0.01 & $-$ \\ \hline
		TiO$_{2}$ & 2.19 $\pm $\ 0.42 & $-$ \\ \hline
		LOI* & 1.93 $\pm $\ 0.81 & 0.30 \\ \hline
	\end{tabular}
	\caption{Average major oxide compositions of basalt rock \citep{wirth1997chengwatana} and felsdpar powder used to produce quenched glass samples.
	*LOI = loss on ignition. \label{TableCompositions}}
\end{table}
\end{center}

\begin{figure}[htb!]
\includegraphics[width=\columnwidth]{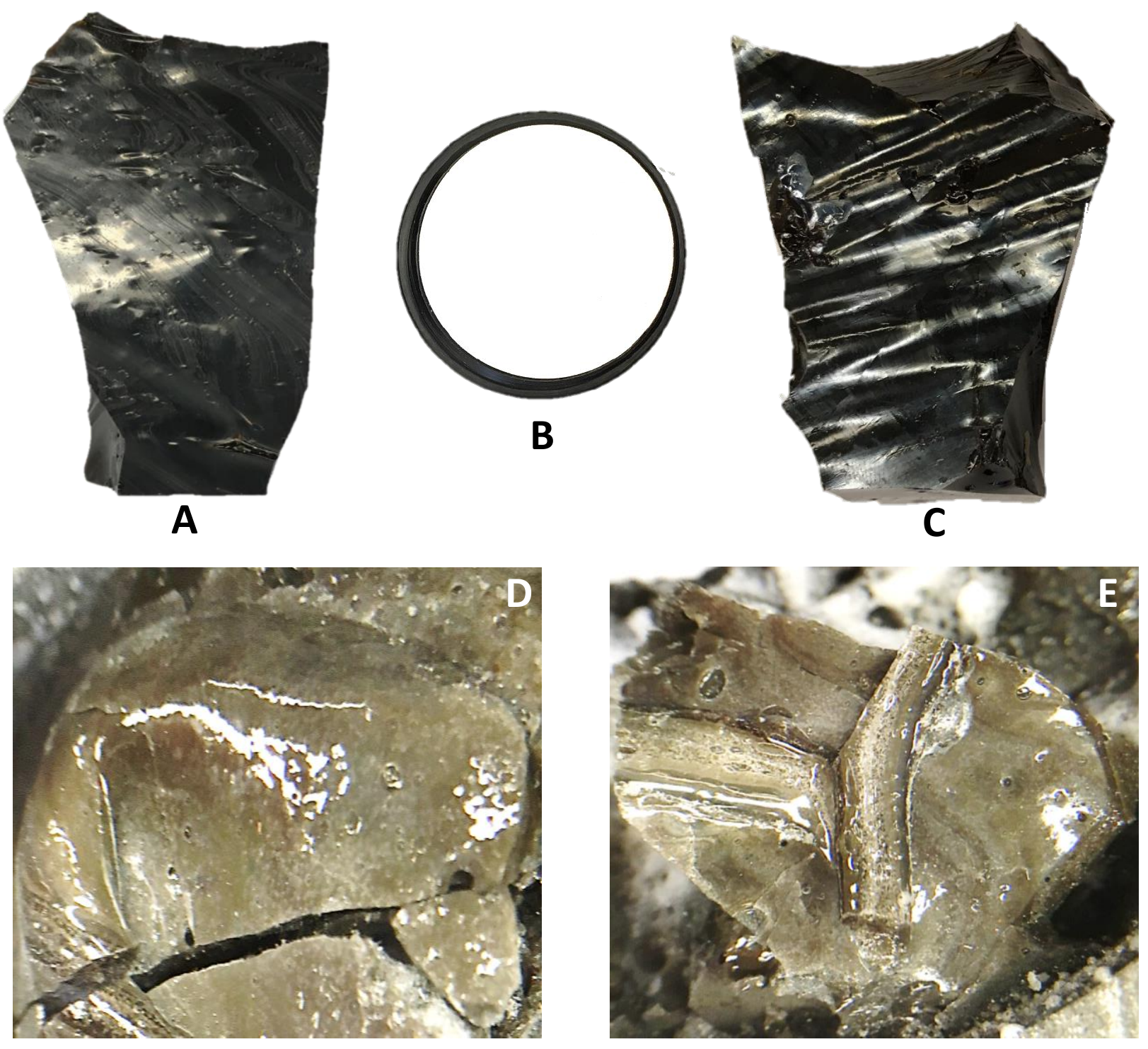}
\caption{Quenched glass samples used to measure reflection. A)\ Smooth basalt glass. B)\ 99\% white reflectance standard. C) Rough basalt glass. D) Smooth feldspar glass. E) Rough feldspar glass. \label{fig2}}
\end{figure}

The basalt rock was melted in a blast furnace at $\sim $1200$ ^{\circ }$C at the Syracuse University Lava Project\footnote{http://lavaproject.syr.edu/making-lava/making.html}. The melt was poured onto a dry sand slope outdoors and allowed to cool for $\sim $10 minutes in air. The rough textured basalt quenched glass formed by the melt running down the slope and exposure to wind before the melt solidified, creating the ripples in the glass (Figure \ref{fig2}, C). The smooth glass is the back of the rough glass sample, which was not subaerially exposed (Figure \ref{fig2}, A).

The feldspar powder was heated to $\sim $1300-1500$ ^{\circ }$C in a graphite crucible in an Inductotherm induction furnace\footnote{Merton C. Flemings Materials Processing Laboratory, MIT} for $\sim $30 minutes to produce the molten rock/lava. The viscous feldspar melt was allowed to cool for $\sim $10 minutes in air to produce a smooth textured quenched glass sample (Figure \ref{fig2}, D). The rough textured feldspar quenched glass was created by streams of viscous melt poured over a flat, smooth layer of feldspar melt (Figure \ref{fig2}, E).

\subsection{Experimental Determination of Reflection from Quenched Glasses} \label{experimentmethod}

We conducted reflectance measurements on our quenched glass samples obtained from melting crushed basalt rock and soda feldspar powder (Figure \ref{fig2}). Reflectance measurements were performed on both the rough and smooth glass textures in order to determine whether surface roughness had a quantifiable effect on the reflectivity of the glasses.

The glass samples were illuminated by a 25~W white LED light source, allowing reflection from the samples to be measured across the visible spectrum (400 $-$ 700~nm).

We designed our experimental setup to measure reflection along the specular reflection direction from both compositions of our rough and smooth glass samples. Specular reflection occurs when the angle of incidence of the incoming light is equal to the angle of reflection from the surface of interest, as measured from the surface normal, the incident and reflected directions are in the same plane, and on opposite sides of the surface normal. At secondary eclipse, the incidence angle and reflected angle are equal, but this is not specular reflection, as only the first of the three conditions is met. However, we measured reflectance at the specular reflection angle because it is easiest to measure and gives us an upper limit on the reflectance from the quenched glass samples, assuming that the specular reflection component is stronger than the retroreflective component. We then used the geometric albedo equation from \citet{1975lpsa.book.....S} to get an upper limit on the planetary geometric albedo (Section \ref{modelmethod}, equation (\ref{eqnNewRho})). For more elaborate modeling, we used a Phong specular reflectance model to translate the reflectance measured in the specular direction into the reflectance measured in the direction a telescope would measure the reflectance from an exoplanet by multiplying the reflectance coefficient with a power of the cosine of the angle between the incident ray and the specular ray (see Section \ref{reflsimulations}).

Though our experimental setup was designed to measure reflection along the specular reflection direction, reflection from most surfaces is always a combination of both specular and diffuse reflection, hence, the reflectance value we measure is a combination of specular and diffuse reflection.

We used an ASEQ instruments LR1 broad spectral range spectrometer (300$-$1000~nm) to measure the reflected counts (intensity when calibrated) from quenched glass samples illuminated by the white LED light source. In order to calculate the reflectivity/albedo of a material, the reflected counts from the material must be divided by the reflected counts from a reference standard $-$ a material of known reflectivity $-$ illuminated by the same light source, under the same conditions. This results in a relative reflectivity value for the material. Multiplying the relative reflectivity by the known reflectivity of the reference standard yields the absolute reflectivity value of the material. We used a 99\% white reflectance standard as our reference standard  (Figure \ref{fig2}, B).

\begin{figure}[htb!]
\includegraphics[width=\columnwidth]{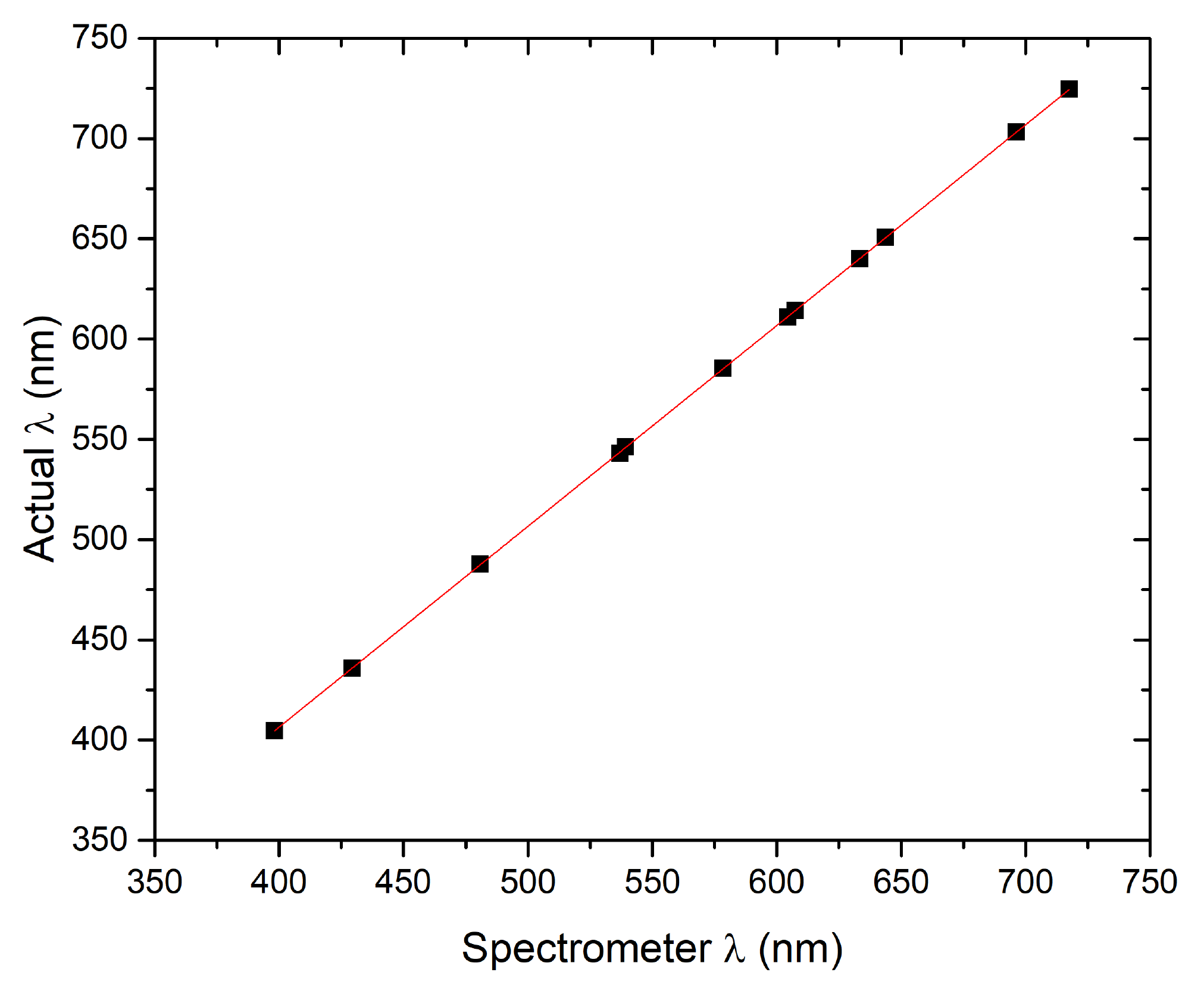}
\caption{Wavelength calibration of the spectrometer across the visible spectrum to account for wavelength shifting. $\protect\lambda _{true}$ is the true/actual wavelength of the spectral line, and $\protect\lambda _{spectrometer}$ is the wavelength of the line identified by the spectrometer. Error bars on data points are smaller than symbols. The red line indicates an even spectral shift across the visible wavelength range. The resulting linear fit to the data (equation (\protect\ref{eqn3})) is used to calibrate the wavelength scale of the spectrometer. Calibration is required to compensate for the systematic error (wavelength shift) introduced by the spectrometer. \label{fig3}}
\end{figure}

We calibrated the wavelength scale of the spectrometer in order to identify how the spectrometer shifts spectral lines. We used a fluorescent ceiling lamp, a TMN-2 crater point vacuum tube, and an INS-1 neon cold cathode vacuum tube lamp to cover the visible wavelength range for calibration. Using the NIST Atomic Spectra Database\footnote{https://www.nist.gov/pml/atomic-spectra-database}, we identified the true wavelengths, $\lambda _{true}$, of the highest intensity spectral lines from the calibration sources with those wavelengths identified by the spectrometer, $\lambda _{spectrometer}$ (Figure \ref{fig3}). We fit a linear function (1$\sigma $ uncertainties on parameter values) to the wavelength shift by the spectrometer:

\begin{multline}
\lambda _{true}=5.83111(\pm 0.48244) + \\ 1.00148(\pm 8.31669\times 10^{-4})\cdot\lambda _{spectrometer}\text{ .}  \label{eqn3}
\end{multline}

\begin{figure*}[htb!]
\plotone{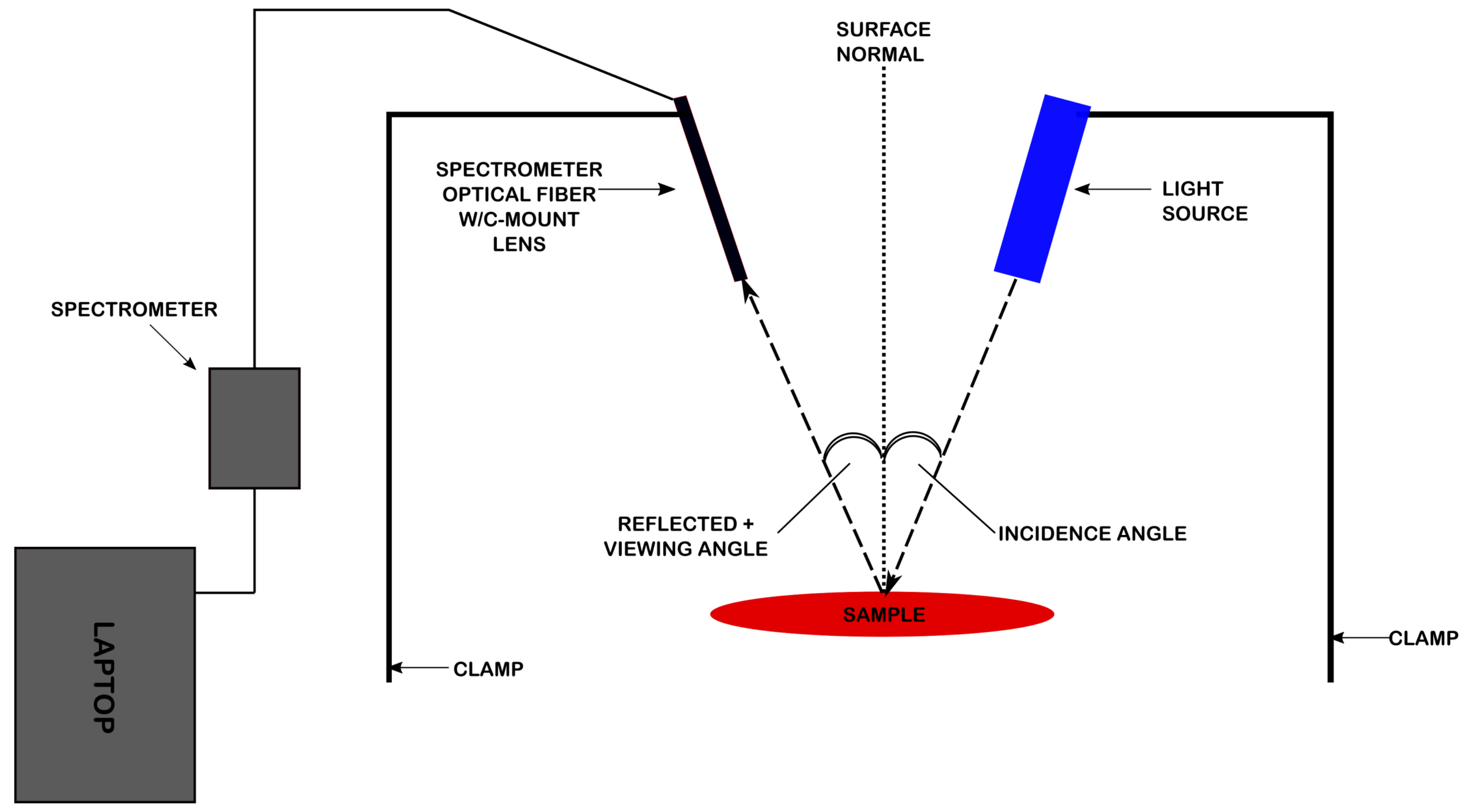}
\caption{Schematic of experimental setup to measure reflection from quenched glass samples. The incidence, reflected, and viewing angles are equal with respect to the surface normal in order to measure reflectance from the quenched glass samples along the specular reflection direction. \label{fig4}}
\vspace*{5mm}
\end{figure*}

The spectrometer optical fiber was attached to a 6-15~mm 1/3" F1.4 C-mount varifocal lens to spatially limit the reflected light entering the fiber. We focused the fiber collecting area on the glass samples illuminated by the center of the light beam for maximum incident intensity. The spectrometer fiber and light source were clamped and attached to support stands 45 cm above the work surface (Figure \ref{fig4}). The clamps can be rotated 360$^{\circ }$.

The spectrometer fiber and light source were clamped at the same angle with respect to the surface normal, so that the angle of incidence was equal to the viewing angle, in order to measure reflected light along the specular reflection direction. The incidence angle (and viewing angle) were varied from 0$^{\circ }$ to 60$^{\circ }$ in 5$^{\circ }$ increments, and reflected counts were measured at each angle for the rough glasses, smooth glasses and reference standard. The experiments were repeated three times.

The wavelength calibration was first applied to the reference standard and quenched glass data. The reflected counts were binned in 10~nm intervals in order to reduce the noise in the data. The reflected counts from the glass samples were divided by the reflected counts from the reference standard at the corresponding incidence angle and the reflectance/reflection coefficient value (Section \ref{modelmethod}) was calculated for each incidence angle. Finally, the spectrally-resolved reflectance measurements were averaged across the visible wavelength range (400$-$700~nm) corresponding to the wavelength range of the white LED light source, and to compare to inferred albedo values from exoplanet observations (see Appendix). We assumed that the incidence angle, reflected angle and viewing angle were all equal, and plot average reflectance as a function of the cosine of the reflected angle (Figures \ref{fig6}$-$\ref{fig10}).

\subsection{Model} \label{modelmethod}

To find the total energy emerging from the exoplanet in the direction of the
observer, we integrate over the visible portion of the planet - the dayside hemisphere - such that:

\begin{equation}
H(\alpha )=\int_{\alpha -\frac{\pi }{2}}^{\frac{\pi }{2}}\cos (\alpha
-\omega )\cos (\omega )d\omega \int_{0}^{\frac{\pi }{2}}\varrho (\eta ,\zeta
,\varphi )\cos ^{3}(\psi )d\psi \text{,}  \label{eqn6}
\end{equation}

\noindent where $\varrho (\eta ,\zeta ,\varphi )$ is the reflection coefficient, $\eta $\ is the cosine of the reflected angle, $\zeta $\ is the cosine of the incidence angle, $\varphi $ is the azimuthal difference between the reflected and incident rays in the local horizontal plane, $\alpha$ is the phase angle, and $\omega $ and $\psi $ are the longitude and latitude on the exoplanet, respectively \citep{1975lpsa.book.....S}. 

If we consider the exoplanet to reflect light according to Lambert's law, $\varrho (\eta ,\zeta ,\varphi )=$ constant, and at secondary eclipse i.e. $\alpha =0$ (assuming the planet is directly behind the star):

\begin{equation}
\phi (\alpha )=\frac{H(\alpha )}{H(0)}=\frac{(\pi -\alpha )\cos (\alpha
)+\sin (\alpha )}{\pi }\text{,}  \label{eqn7}
\end{equation}

\noindent where $\phi (\alpha )$ is the phase function \citep{1975lpsa.book.....S}.

The geometric albedo, $A_{g}$, is defined as the ratio of the planetary flux
at secondary eclipse to the flux from a Lambert disk of the same cross
sectional area. It follows that:

\begin{equation}
A_{g}=\frac{2}{\pi }H(0).  \label{eqn8}
\end{equation}

At secondary eclipse ($\alpha =0$), the reflected angle is equal to the
incidence angle ($\eta =\zeta $), and $\varphi =\pi $. $A_{g}$ can then be
written as:

\begin{equation}
A_{g}=\frac{2}{\pi }\int_{-\frac{\pi }{2}}^{\frac{\pi }{2}}\cos ^{2}(\omega
)d\omega \int_{0}^{\frac{\pi }{2}}\varrho (\eta ,\eta ,\pi )\cos ^{3}(\psi
)d\psi \text{ .}  \label{eqn9}
\end{equation}

Rewriting equation (\ref{eqn9}) in terms of the cosine of the reflected angle, $\eta $, yields \citep{1975lpsa.book.....S}:

\begin{equation}
A_{g}=2\int_{0}^{1}\varrho (\eta ,\eta ,\pi )\eta ^{2}d\eta \text{ .}
\label{eqn10}
\end{equation}

In our experiments, we aligned our viewing angle in the direction of specular reflection and measure $\varrho (\eta ,\eta , 0 )$. We approximate $\varrho (\eta ,\eta ,\pi )$ by $\varrho (\eta ,\eta , 0 )$, and calculate the geometric albedo as:

\begin{equation}
A_{g}=2\int_{0}^{1}\varrho (\eta ,\eta ,0)\eta ^{2}d\eta \text{ .}
\label{eqnNewRho}
\end{equation}

We experimentally measure $\varrho (\eta ,\eta , 0)$ for our quenched glass samples, assuming the incidence angle, reflected angle and viewing angle are all equal. We fit a model to the data to obtain an equation for $\varrho $ as a function of $\eta $. We integrate the experimentally determined function for $\varrho $ over all latitudes and longitudes i.e. all reflected angles on
the dayside hemisphere of the planet (equation (\ref{eqnNewRho})) to obtain a value for $A_{g}$ due to rough and smooth, basalt and feldspar quenched glasses.

\begin{figure}[htb!]
\includegraphics[width=\columnwidth]{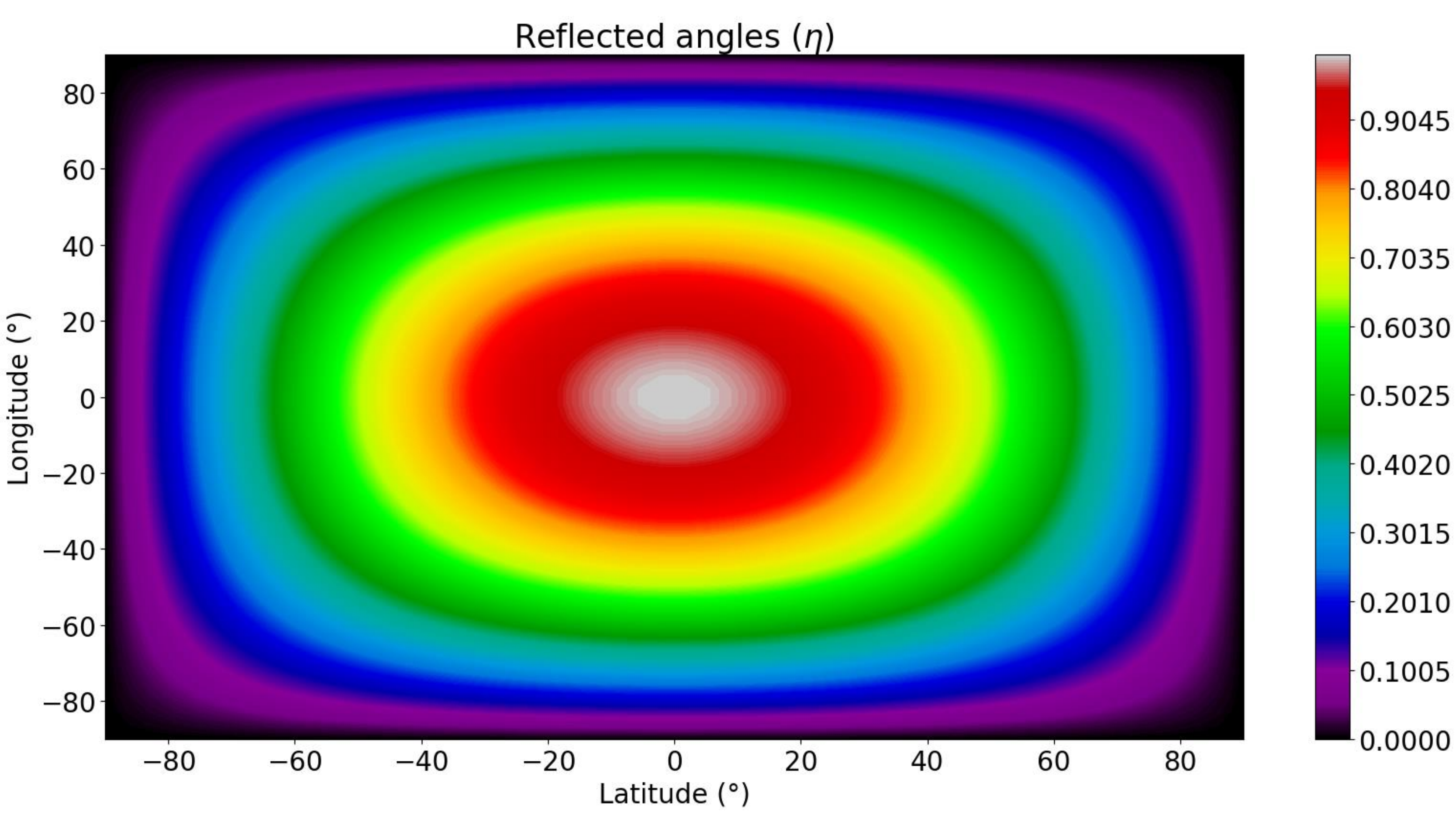}
\caption{Reflected angles ($\eta$) at secondary eclipse as a function of latitude and longitude on the dayside hemisphere of a planet. The color-bar represents the reflected angle and shows the greatest reflected light contribution comes from the center of the planetary disk for $-$20$^{\circ }<$ latitude, longitude $<~$20$^{\circ }.$ \label{fig5}}
\end{figure}

In order to calculate the geometric albedo of a planet with a combination of
lava and quenched glass on the surface, we surveyed non-crystalline solids
literature for specular reflectance values from molten silicates as a proxy
for specular reflectance values for lava. Specular reflection values from molten silicates were measured at an incidence angle of 0$^{\circ }$ and have an average reflectance of 0.15 for compositions with major oxide proportions most similar to our basalt quenched glass sample's composition \citep{zebger2005ultraviolet}. The planet geometric albedo was calculated by varying the amounts of lava and quenched glass on the surface according to:

\begin{equation}
A_{g}=2\int_{0}^{x}\varrho _{glass}(\eta ,\eta ,0 )\eta ^{2}d\eta \text{ }%
+2\int_{x}^{1}\varrho _{lava}(\eta ,\eta ,0 )\eta ^{2}d\eta\text{,}
\label{eqnlavaglass}
\end{equation}%

\noindent where $x$ represents the extent/amount of lava or glass on the surface based on regions of $\eta$ on the planet surface (Figure \ref{fig5}), $\varrho _{glass}(\eta ,\eta ,0 )$ is our experimentally determined reflection coefficient function for different quenched glasses, and $\varrho_{lava}(\eta ,\eta ,0 )=$ 0.15 is the constant reflection coefficient for lava from non-crystalline solids literature. The integral is structured such that we assume lava originates at the substellar point and quenches to glass as the distance increases radially outward from the substellar point, and the quenched glass extends to the poles. When $x=0$, there is an all lava dayside surface, and when $x=1$, there is an all quenched glass (basalt or feldspar, rough or smooth) dayside surface.

We assume plane-parallel rays of incident light for lava-ocean planets however, for extremely close-in exoplanets like Kepler-10 b and CoRoT-7 b, incident light from the host star is in the finite angular size regime. 

For plane-parallel incident light, the exoplanet is illuminated evenly into a dayside that comprises one hemisphere (50\% of the planet surface) and a nightside. For extremely close-in exoplanets, the finite angular size of the star determines ``illumination zones" on the planet surface $-$ a fully-illuminated zone and a penumbral zone on the dayside, and a nightside. The dayside of the planet will be $>$ 50\% of the surface, and different reflected luminosities will be observed for the fully-illuminated zone and penumbral zone \citep{carter2019estimation}.

The difference in reflected luminosity between the fully-illuminated zone and penumbral zone is $\sim $18\% ($\sim $14~ppm) at secondary eclipse \citep{carter2019estimation}. The maximum photometric precision of \textit{Kepler} and \textit{TESS} are approximately 29~ppm and 60~ppm respectively \citep{gilliland2011kepler,ricker2014transiting}, hence the luminosity difference between the plane-parallel and finite angular size regimes is not detectable with current telescopes.

\section{Results} \label{sec:results}

We find that reflection from samples of rough and smooth quenched glass results in low planetary geometric albedos: $\lesssim$~0.09 for basalt quenched glass, and $\lesssim$~0.02 for feldspar quenched glass. Using specular reflectance values from molten silicates as a proxy for specular reflectance values for lava also results in low planetary geometric albedos ($\lesssim$~0.1).

\subsection{Planetary Albedo: Quenched Glass} \label{albedoglassresults}

We consider two end-member quenched glass compositions derived from basalt (mafic) and feldspar (felsic) lavas. We find the albedo of a planet dayside surface covered entirely in rough or smooth basalt quenched glass is $\sim $0.09, and the albedo of a planet dayside surface covered entirely in rough or smooth feldspar quenched glass is $\sim $0.02.

The quenched glass reflectance data, measured along the specular reflection direction, are a natural combination of specular and diffuse reflection. The absolute specularity of the quenched glasses cannot be determined from the data because the angle between the specularly reflected ray and viewing angle is unconstrained. We model specular reflection and attempt to quantify the specularity of the quenched glasses in Section~\ref{reflsimulations}.

Reflection from quenched glasses has the greatest contribution to reflectance at small incidence angles ($<$~20$^{\circ }$) and has a negligible contribution when considered over the entire planet, even if the observer is aligned with the specular reflection direction. We  find  that  reflectance  does  not  vary  significantly  as  a  function  of  wavelength  across  the  visible wavelength  range, the shape of the reflectance curves do not change appreciably with incidence angle, and the amplitude of the reflectance curve decreases with increasing incidence angle (Figure \ref{ReflWavApx}).

Surface roughness does not change our results away from a low planetary albedo. The fact that specular reflection from a rough surface scatters light into a cone with some angular extent, as compared to a smooth surface that scatters light in a single direction, does not increase the albedo. It is worth noting, however, that for our basalt quenched glass samples, there is a more rapid decrease in reflectance values for the smooth glass sample (Figure \ref{fig8}) and a more gradual decrease in reflectance values for the rough glass sample (Figure \ref{fig6}), because reflection from the rough glass sample can be seen at more angles.

\subsubsection{Basalt Quenched Glass}

We find the albedo of a planet dayside surface covered entirely in rough basalt quenched glass is $0.09_{-\text{0.01}}^{+\text{0.02}}$, and smooth basalt quenched glass is $0.09_{-\text{0.01}}^{+\text{0.03}}$. The calculated albedo values are from the model (Section~\ref{modelmethod}, equation~(\ref{eqnNewRho})), with our measured experimental data as input.

The measured experimental data are fitted by an exponential model of the form $e^{R(x+A)}+C$. An exponential model was chosen to account for the highly directional nature of specular reflection. Specular reflection can only be observed within a narrow range of angles around 0$^{\circ }$ and decays rapidly away from normal incidence, which agrees well with an exponential model. The model fit parameters and 1$\sigma$ uncertainties are shown in Table \ref{Table2}. The resulting planetary geometric albedo was calculated by integrating the exponential model fit to the basalt quenched glass reflectance data over all reflected angles according to equation (\ref{eqnNewRho}).

\begin{table}[htb!]\centering
	
	\begin{tabular}{|c|c|c|c|c|c|}
		\hline
		Basalt & R & A & C \\ \hline
		Rough & 19.60 $\pm $\ 4.20 & -1.03 $\pm $\ 0.01 & 0.066 $\pm $\ 0.011 \\ \hline
		Smooth & 35.39 $\pm $\ 6.06 & -1.008 $\pm $\ 0.006 & 0.080 $\pm $\ 0.022 \\ \hline
	\end{tabular}
	\caption{Exponential model fit parameter values (1$\sigma$ uncertainties) for rough and smooth basalt quenched glass reflectance data. \label{Table2}}
	
\end{table}

\begin{figure}[t!]
\includegraphics[width=\columnwidth]{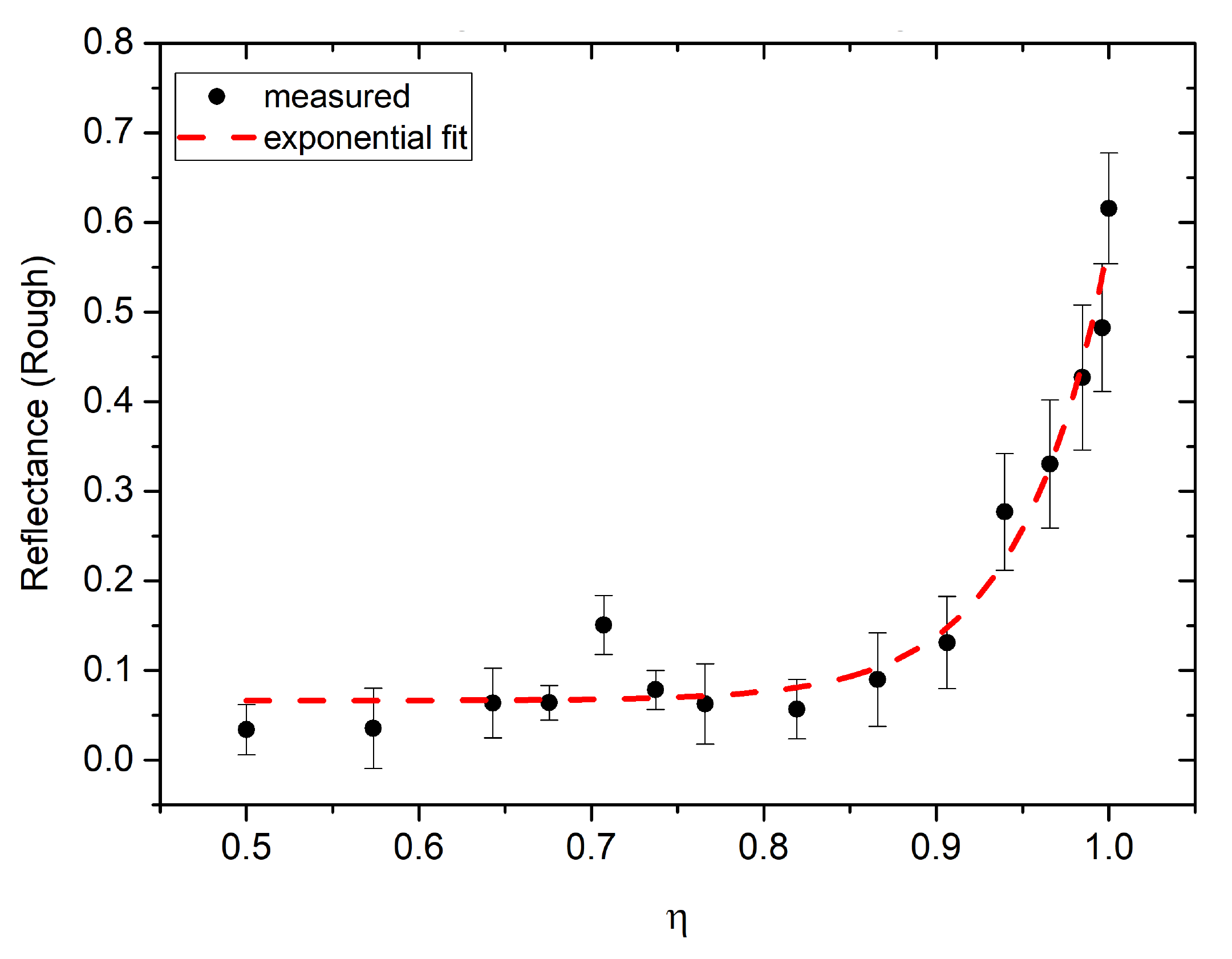}
\caption{Reflectance vs. $\eta$ data of rough basalt glass sample for incidence angles 0${{}^\circ}$-60${{}^\circ}$. The black points are the experimentally measured data with 1$\protect\sigma $ errorbars, and the red dashed line is the exponential model fit to data. Reflectance values are large for small reflected angles ($<$ 20$^{\circ }$) and decrease rapidly with increasing reflected angle. Increase in reflectance at 45$^{\circ }$ is a real feature which currently does not have a definitive explanation.\label{fig6}}
\end{figure}

\begin{figure}[b!]
\includegraphics[width=\columnwidth]{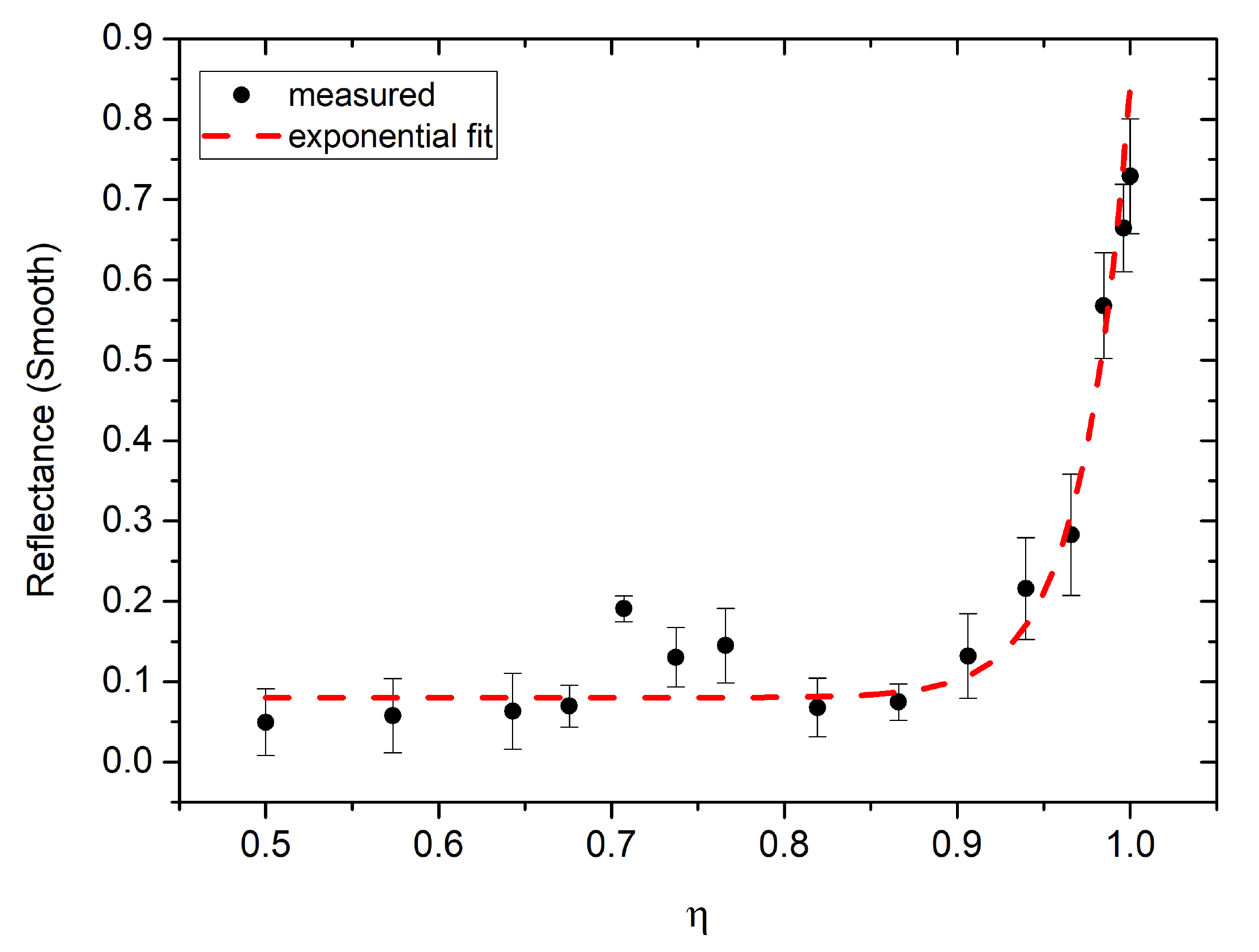}
\caption{Reflectance vs. $\eta$ data of smooth basalt glass sample for incidence angles 0${{}^\circ}$-60${{}^\circ}$. The black points are the experimentally measured data with 1$\protect\sigma $ errorbars, and the red dashed line is the exponential model fit to data. Reflectance values are large for small reflected angles ($<$ 15$^{\circ }$) and decrease rapidly with increasing reflected angle. Increase in reflectance at 45$^{\circ }$ is a real feature which currently does not have a definitive explanation. \label{fig8}}
\end{figure}

Reflectance vs. $\eta $ data (incidence angles 0$^{\circ }-$60$^{\circ}$) of the rough basalt glass sample and smooth basalt glass sample are presented in Figure \ref{fig6} and Figure \ref{fig8}, respectively\footnote{When measuring reflectance along the specular reflection direction, the incidence angle and reflected angle are equal so incidence angle ($\arccos{\zeta}$) and reflected angle ($\arccos{\eta}$) can be used interchangeably.}.

Notably for both our rough and smooth basalt quenched glass samples, there is a small but significant increase in reflectance at  45$^{\circ }$ (Figures~\ref{fig6}, \ref{fig8}). The feature is present after multiple measurements and appears to be real. We have no conclusive explanation for the feature but are exploring possibilities related to off-specular reflection due to surface roughness \citep{torrance1967theory}, or microscopic crystalline structure features in the samples.

\subsubsection{Feldspar Quenched Glass}

We find the albedo of a planet dayside surface covered entirely in rough feldspar quenched glass is $0.02_{-\text{0.01}}^{+\text{0.02}}$, and smooth feldspar quenched glass is $0.02 \pm 0.01$. The calculated albedo values are from the model (Section~\ref{modelmethod}, equation~(\ref{eqnNewRho})), with our measured experimental data as input.

The measured experimental data are fitted by an exponential model of the form $e^{R(x+A)}+C$. The model fit parameters and 1$\sigma$ uncertainties are shown in Table \ref{Table3}. The resulting planetary geometric albedo was calculated by integrating the exponential model fit to the feldspar quenched glass reflectance data over all reflected angles according to equation (\ref{eqnNewRho}).

\begin{table}[htb!]\centering
	
	\begin{tabular}{|c|c|c|c|c|c|}
		\hline
		Feldspar & R & A & C \\ \hline
		Rough & 8.47 $\pm $\ 1.05 & -1.29 $\pm $\ 0.035 & 0.010  $\pm $\ 0.002 \\ \hline
		Smooth & 10.12 $\pm $\ 1.32 & -1.24 $\pm $\ 0.030 & 0.014 $\pm $\ 0.003 \\ \hline
	\end{tabular}
	\caption{Exponential model fit parameter values (1$\sigma$ uncertainties) for rough and smooth feldspar quenched glass reflectance data. \label{Table3}}
	
\end{table}

\begin{figure}[htb!]
\includegraphics[width=\columnwidth]{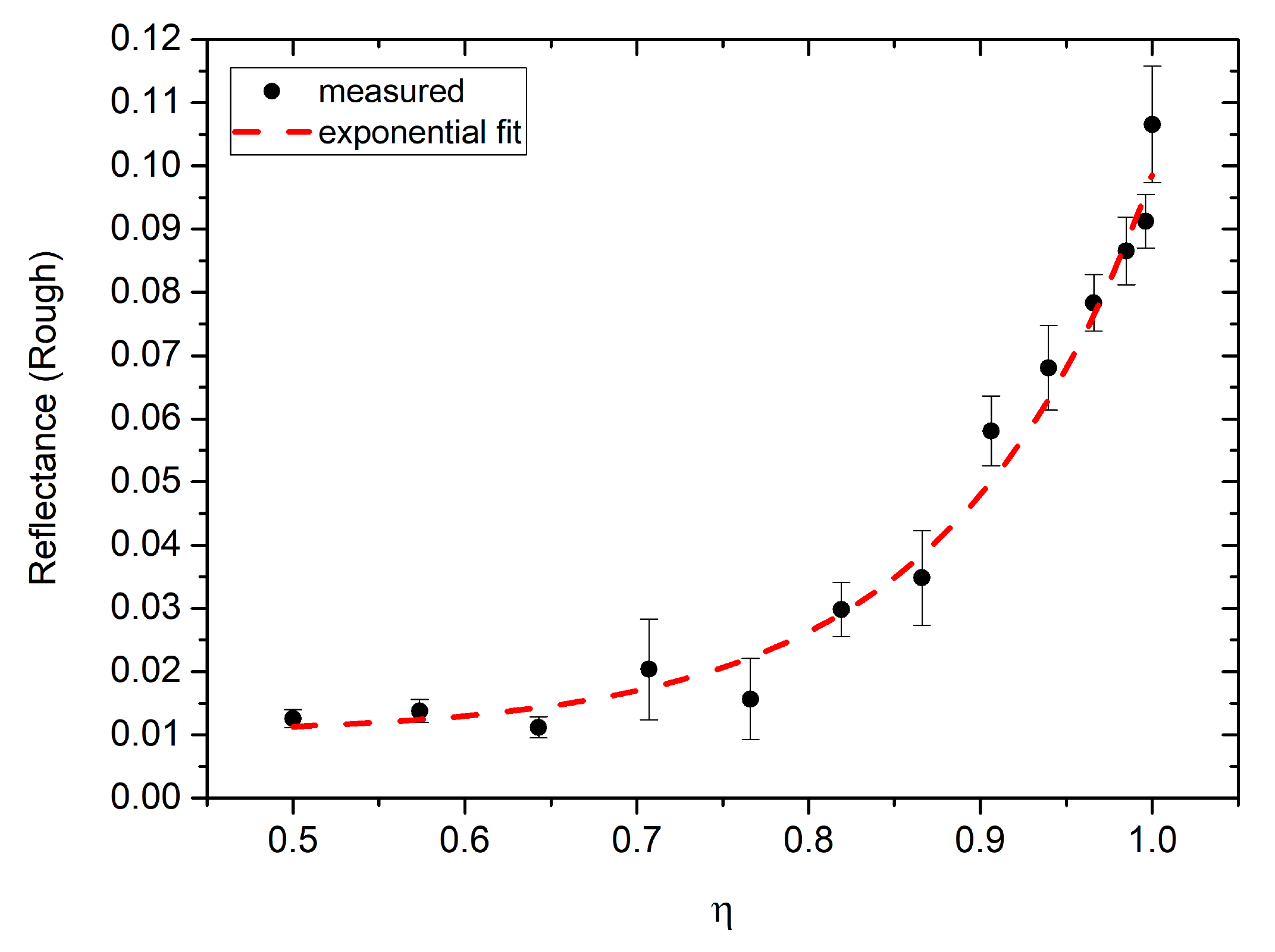}
\caption{Reflectance vs. $\eta$ data of rough feldspar glass sample for incidence angles 0${{}^\circ}$-60${{}^\circ}$. The black points are the experimentally measured data with 1$\protect\sigma $ errorbars, and the red dashed line is the exponential model fit to data. Reflectance values decrease with increasing reflected angle. \label{fig9}}
\end{figure}

\begin{figure}[htb!]
\includegraphics[width=\columnwidth]{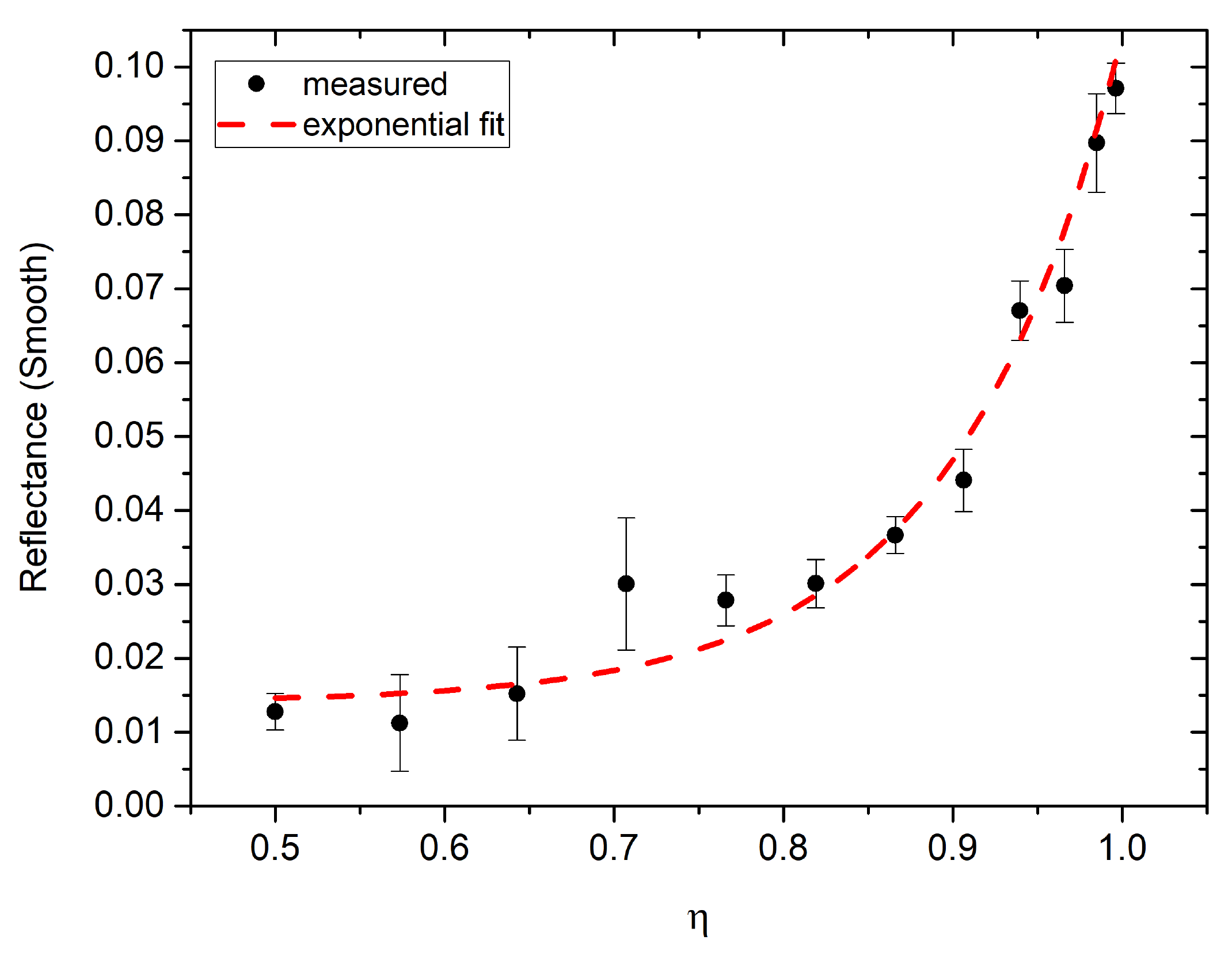}
\caption{Reflectance vs. $\eta$ data of smooth feldspar glass sample for incidence angles 0${{}^\circ}$-60${{}^\circ}$. The black points are the experimentally measured data with 1$\protect\sigma $ errorbars, and the red dashed line is the exponential model fit to data. Reflectance values decrease with increasing reflected angle. \label{fig10}}
\end{figure}

Reflectance vs. $\eta $ data (incidence angles 0$^{\circ }-$60$^{\circ}$) of the rough feldspar glass sample and smooth feldspar glass sample are presented in Figure \ref{fig9} and Figure \ref{fig10}, respectively.

The feldspar reflectance values are consistently low and the decrease in the reflectance values is gradual and inconsistent with the highly directional nature expected of specular reflection, where reflectance values decay rapidly away from normal incidence, as seen in the basalt quenched glass reflectance data (Figures \ref{fig6}, \ref{fig8}).

\bigskip
Both specular and diffuse reflection from quenched glasses, independent of glass composition, results in low planetary geometric albedos. We focus on describing reflectance results along the specular reflection direction, and measurements of diffuse reflection from quenched glasses of different cooled molten silicate compositions in other studies found low reflectance values ($\lesssim$~0.1) at visible wavelengths as well \citep{zebger2005ultraviolet, nowack2001ultraviolet}.

\subsection{Planetary Albedo: Lava and Quenched Glass} \label{albedoglasslavaresults}

Lava-ocean planets may have a combination of lava and quenched glass on their surfaces. Here we consider a planet covered in lava as an extreme end-member, as well as mixtures where lava is present near the substellar point and basalt/feldspar quenched glass is present beyond. We find that no matter the surface combination of lava and quenched glass, the dayside geometric albedo is always low, $A_{g} \lesssim$ 0.1.

\begin{figure}[htb!]
\includegraphics[width=\columnwidth]{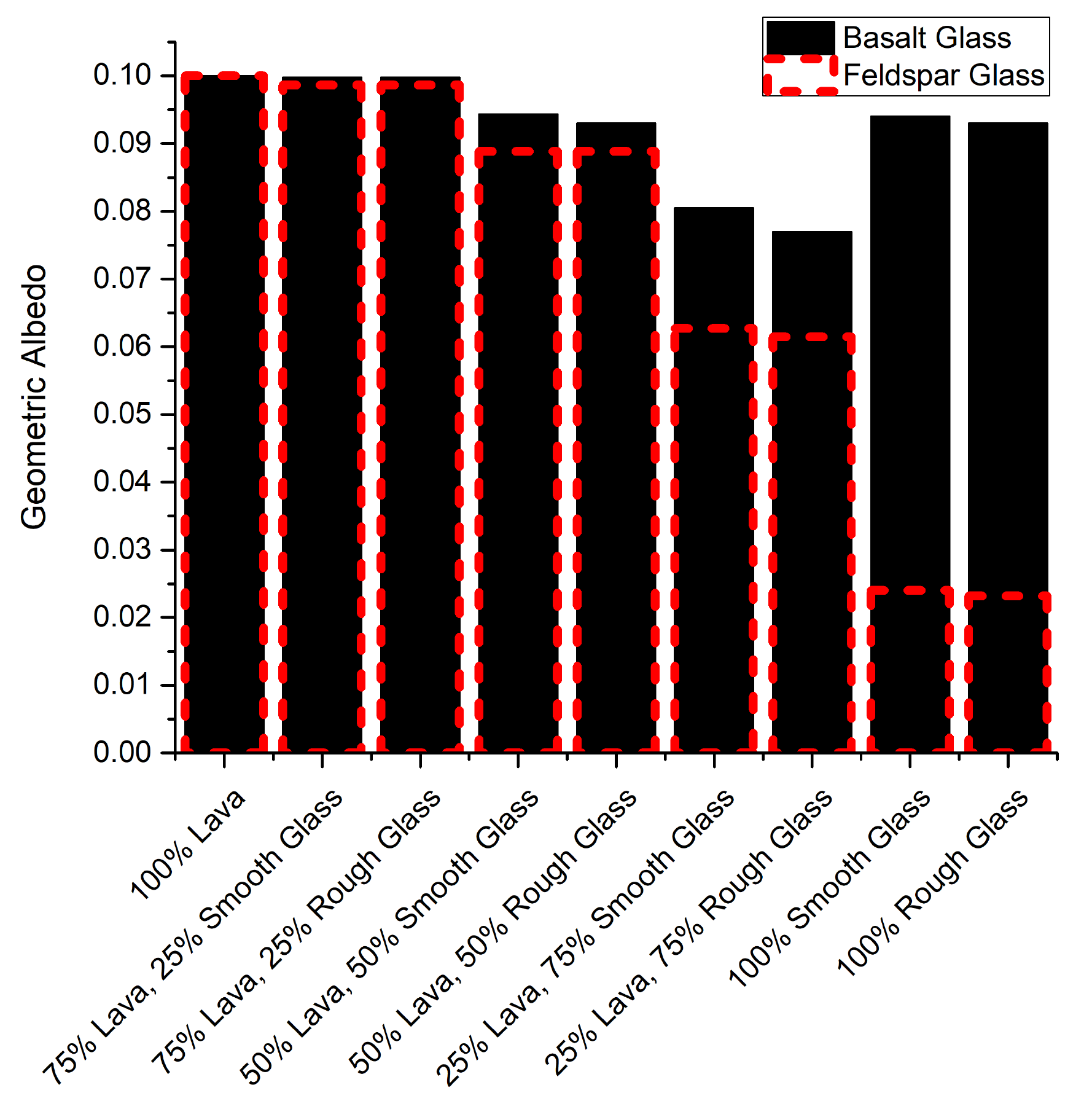}
\caption{Planetary geometric albedo for a combination of lava and quenched glass on the surface. The upper limit of the dayside geometric albedo is $\sim$0.1 Basalt quenched glass albedo values are represented by black bars and feldspar quenched glass albedo values are represented by red dashed line bars. The amount/extent of the lava and glass on the surface is determined by the integral bound ($x$) on the reflected angle in equation (\protect\ref{eqnlavaglass}). At $\protect\eta =0$, the surface is all lava and at $\protect\eta =1$, the surface is all rough or smooth quenched glass on the dayside hemisphere.\label{figcomblava}}
\end{figure}

The amount/extent of lava and quenched glass on the surface was varied by varying the integral bound, $x$, on the reflected angle in equation (\ref{eqnlavaglass}). The geometric albedo of a planet with a combination of lava and quenched glass (rough or smooth) on the surface is shown in Figure \ref{figcomblava}.

For a planet with a dayside hemisphere completely covered in lava, we find $A_{g} \lesssim$ 0.1.

The albedo value for an all-lava planet was obtained from non-crystalline solids literature. \citet{zebger2005ultraviolet} measured specular reflection from molten silicates, and varied the amounts of CaO, Fe$_2$O$_3$ and SiO$_2$ to create different mixture compositions. We selected a molten silicate composition from \citet{zebger2005ultraviolet} that was sufficiently similar to our basalt quenched glass composition as a proxy for lava. There were no feldspar equivalent molten silicate compositions so we conservatively assume the same reflectance value for both basalt and feldspar lavas.

The specular albedos are low ($<$ 0.18) for all the molten silicate compositions considered in \citet{zebger2005ultraviolet} at the measured VIS wavelengths. UV/VIS specular reflection measurements were obtained for the different molten silicate compositions at a single incidence and viewing angle of 0$^{\circ }$.

Since specular reflection from molten silicates was measured at a single angle, there is only one reflectance value for lava, and we had to set the reflection coefficient for lava to be this constant reflectance value ($\varrho_{lava}(\eta ,\eta , 0)=$ 0.15). This effectively gives us an upper limit on the albedo of an all-lava dayside surface because we cannot determine the angular dependence of the reflection coefficient for lava.

\section{Discussion} \label{sec:discussion}

We support our results with simulations to constrain the reflectance from lava worlds, and discuss the implications of our results for lava and quenched glasses as sources of reflected light on lava-ocean exoplanets. We describe potential differences between quenched glasses in this study and quenched glasses formed under lava-ocean exoplanet conditions, other potential high albedo surfaces, and discuss reflection in the atmospheres of hot super Earths. We discuss the challenges involved in measuring reflection from molten lava in the laboratory, and conclude the discussion with possibilities of lava-ocean planet candidates in \textit{TESS} data for future characterization.

\subsection{Verification of Albedo Estimates via Reflectance Simulations} \label{reflsimulations}

We performed simple reflectance simulations using a diffuse reflectance model and a Phong specular reflectance model \citep{phong1975illumination} to verify our experimental quenched glass albedo estimates with regards to the geometric effects arising from extremely close-in lava planets and their host stars (finte angular size regime - Section \ref{modelmethod}).

We find that the difference between assuming plane-parallel incident light and incident light in the finite angular size regime produces a negligible effect on the calculated albedo values. 

The Monte Carlo diffuse reflectance model produced albedos within 5\% of our experimentally determined basalt quenched glass planetary albedo values (Table \ref{TableMC}) by using the same measured reflectance parameters in equation (\ref{eqnNewRho}). The basalt quenched glass reflectance data are not consistent with purely diffuse reflection, and are a combination of specular and diffuse reflection. The basalt glass data can be approximated by a Phong specular reflectance model with an empirical specular falloff  value, $n$, between 1 and 10, and a diffuse reflectance model. Specular reflection from the Phong reflectance model produced albedo values that were 1.6-1.8 times lower than our model values for $1 < n <10$ (Table \ref{TableMC}). The smoother the planet surface (i.e. the higher the specularity), the lower the albedo. 

The feldspar quenched glass data and planetary albedo values can be approximated by a diffuse reflectance model (Table \ref{TableMC}).

\begin{table*}[htb!]
	\begin{tabular}{|c|c|c|c|c|c|}
		\hline
		CoRoT-7 b & Diffuse & Phong n=1 & Phong n=10 & Phong n=100 & Phong n=1000 \\ \hline
		Basalt - Rough & 0.088 & 0.053 & 0.017 & 0.0023 & 0.0002 \\ \hline
		Basalt - Smooth & 0.099 & 0.057 & 0.020 & 0.0030 & 0.0003 \\ \hline
		Feldspar - Rough & 0.024 & 0.014 & 0.0037 & 0.0004 & 0.00005 \\ \hline
		Feldspar - Smooth & 0.024 & 0.013 & 0.0037 & 0.0005 & 0.00005 \\ \hline
	\end{tabular}
	\caption{Geometric albedo values for rough and smooth basalt and feldspar quenched glass surfaces for CoRoT-7 b system parameters, using Monte Carlo ray-tracing simulations. The geometric effects of incident light for extremely close-in planets (finite angular size regime) on the planet albedo are considered. 
	\label{TableMC}}
	
\end{table*}

We used Monte Carlo ray-tracing, and modeled the star as a disk where each point on the star has an equal probability of emitting a ray in the direction of the planet. No limb darkening was used. The observer was considered to be a disk with the same diameter as the planet, situated infinitely far away from the planet, with its surface normal directed towards the planet. 

The quenched glass experimental albedo datasets were interpreted as the flux measured by the observer from the quenched glass sample ($F_{M}$) divided by the flux of a reference standard ($F_{R}$) placed at the same geometry, such that: 

\begin{eqnarray}
A'(\alpha) = \frac{F_{M}(\alpha)}{F_{R}(\alpha)} = \frac{F_{M}(\alpha)}{I\cos{\alpha}}\\
\rightarrow F_{M}(\alpha) = A'(\alpha)I\cos{\alpha}
\label{datatomodel}
\end{eqnarray}

\noindent where $A'(\alpha)$ is the measured reflectance, $\alpha$ is the angle of incidence (angle between the planet surface normal and the ray connecting it with the point on the surface of the star), and $I$ is the incident light intensity. 

We calculated the intensity of each ray reaching the observer based on both a Phong specular reflectance model and a diffuse reflectance model. 

For specular reflectance, we calculate the specular intensity, $F_{S}'$, of a ray from the measured data using the Phong model \citep{phong1975illumination} as:

\begin{equation}
F_{S}'= A'(\alpha)I\cos{\alpha}\cos(\alpha-\beta)^n
\label{specular}
\end{equation}

\noindent where $\beta$ is the observer angle (angle between the observer ray and the surface normal), and $n$ is an empirical parameter specifying specular falloff (the larger the value of n, the smoother/shinier the surface). We varied the values of $n$ (between 1 and 1000) and $\beta$  to constrain the value of $F_{S}'$. 

Rewriting equation (\ref{specular}) in terms of our model parameters in Section \ref{modelmethod} yields:

\begin{equation}
F_{S}'= A'(\alpha)I\eta\cos(\arccos{\eta}-\arccos{\zeta})^n
\label{specularSobolev}
\end{equation}

For diffuse reflectance, we calculated the intensity, $F_{L}'$, as:

\begin{eqnarray}
F_{L}'= A'(\alpha)I\cos{\alpha}\\
F_{L}'= A'(\alpha)I\eta.
\label{lambertian}
\end{eqnarray}

\subsection{Difference in Reflectivity of Basalt and Feldspar Quenched Glasses} \label{basaltfeldsparglasses}

Our feldspar quenched glass had lower albedos than our basalt quenched glass by $\sim $7\%. The difference in reflectivity between the glasses can be attributed to the compositional differences between the materials used to produce the glasses. Basalt and feldspar rock powders were chosen to represent end-member mafic and felsic lavas respectively.

A possible explanation for the low reflectivity of our feldspar quenched glasses is the formation of the quenched glass crust. The viscosity of the melt/lava affects the formation of quenched glass. Felsic materials are silica rich ($>$ 65\% by weight), have lower melting points ($\sim $1100 K) compared to silica poor/mafic materials ($<$ 55\% by weight; $\sim $1500 K melting point), and form highly viscous melts. Felsic materials are also light in color as compared to mafic materials due to their enrichment in potassium, sodium and aluminum (mafic materials are enriched in iron, calcium and magnesium) \citep{pidwirny2006characteristics}.
 
The felsdpar powder used in our experiment resulted in a viscous melt containing large gas bubbles which disrupted the initial quenched glass crust that had formed by forcing molten material below to push through the crust. The disruption of the quenched glass crust formed over the feldspar melt could have slowed melt cooling, which may have resulted in partial solidification of the melt through crystallization rather than quenching \citep{griffiths2000dynamics}. Crystallization affects surface roughness and could have reduced the reflectivity of the feldspar quenched glass.

The bubbles present in the feldspar quenched glass likely formed from H$_{2}$O, CO$_{2}$ or other organic gases present in the starting material.

\subsection{Earth Experiments vs. Exoplanet Conditions} \label{expconditions}

The quenched glasses produced in this study were at Earth's surface pressures and atmospheric composition, and it is worth noting how quenched glasses formed under conditions on lava-ocean exoplanets may differ. 

Lava-ocean exoplanets are expected to have low pressure atmospheres and hence low oxygen partial pressures (p$_{\rm O_2}$). While volatile elements such as H, C, and S are expected to be lost from the atmosphere, mineral vapor atmospheres consisting of Na, K, O$_{2}$, O, SiO etc. have been theorized on hot super Earths \citep{schaefer2009chemistry, ito2015theoretical}. 

\citet{cannon2017spectral} showed that quenched glass composition and p$_{\rm O_2}$ formation conditions strongly affect reflectance. At higher p$_{\rm O_2}$, Fe$^{3+}$ content increases and reflectance decreases due to the Fe$^{2+}$-Fe$^{3+}$ charge-transfer band ($\sim$600 nm) absorption strength increasing \citep{nowack2001ultraviolet, zebger2005ultraviolet, cannon2017spectral}. An increase in log(p$_{\rm O_2}$) by 4 units can decrease glass reflectance values by $\sim$15\% at visible wavelengths \citep{cannon2017spectral}. Composition differences between glasses, particularly transition metal content (Fe and Ti have strong absorption features), can increase/decrease reflectance values by 10-50\% \citep{cannon2017spectral}.

Although a small subset of glasses from \citet{cannon2017spectral} formed under extremely low oxygen conditions and containing little/no iron had high reflectance values, the glasses were synthetic and the conditions and composition of the synthetic glasses are unlikely to be the same on hot super Earth exoplanets when factoring in planet size, planet formation conditions etc. \citep{kite2016atmosphere}. Without observed spectra of hot super Earth atmospheres/surfaces, we cannot rule out a lack of oxygen in the atmosphere or iron on the surface. \citet{cannon2017spectral} notes that efforts should be made to use realistic glasses for a given planetary body, which we have attempted to do by using naturally occurring materials to create the quenched glasses in this study.

The reflectance values of molten silicates from non-crystalline solids literature \citep{nowack2001ultraviolet, zebger2005ultraviolet} used as proxies for molten lava have substantial Fe$_{2}$O$_{3}$ content, and lavas with lower/higher total oxidized Fe content will have different reflectance values than the average value used in this study for molten silicates.

Earth's atmosphere is highly oxidizing and basaltic quenched glasses produced in air can have Fe$^{3+}$ values as high as 60\% \citep{cannon2017spectral}. In contrast, quenched glasses formed in highly reducing atmospheres e.g. the Moon or Mercury, contain little or no Fe$^{3+}$ \citep{cannon2017spectral}. Hot super Earth p$_{\rm O_2}$ levels can vary depending on the initial Fe$^{2+}$/Fe$^{3+}$ ratio of the planet, and atmospheric escape which can increase or decrease p$_{\rm O_2}$ levels depending on whether oxygen is concentrated in or removed from the atmosphere. Quenched glasses formed under lower p$_{\rm O_2}$ conditions (hence with lower Fe$^{3+}$ content) on hot super Earths could have higher reflectance values than the quenched glasses produced in this study. Though the Fe$^{2+}$/Fe$^{3+}$ ratio of hot super Earths is unknown, it would indeed affect the reflectivity of quenched glasses formed on hot super Earths, producing glasses with higher/lower reflectance values than the quenched glasses in this study.

While quenched glasses formed on hot super Earths may have higher reflectance values than the quenched glasses in this study, we expect a $<$15-20\% increase in reflectance values which will still result in low albedos ($A_{g} <$  0.22 for basalt quenched glass and $A_{g} <$ 0.16 for feldspar quenched glass) and not change the overall result of this study.

\subsection{Evolved High Albedo Surfaces} \label{brightsurfaces}

We have shown that basalt and feldspar surfaces of molten lava or quenched glass have low albedos.  Beyond these mafic and felsic surface compositions, there are a specific but narrow set of initial conditions that may lead to evolved high albedo surfaces composed of Ca/Al oxides \citep{kite2016atmosphere}. For completeness, we present a summary of the concept.

A hot super Earth's surface may evolve due to fractional vaporization of melt on the surface, with winds removing the volatiles to the planet night side or out to space.  As the more easily vaporized material is stripped away, less volatile but still molten crustal material remains. The final evolved surface composition results from a competition between the rate at which the volatiles are removed from the dayside of planet compared to the rate of mass recycling between the melt pool and solid interior \citep{kite2016atmosphere}.

\citet{kite2016atmosphere} finds a scenario that may lead to a high albedo molten surface. For substellar temperatures $>$ 2400 K, atmospheric transport dominates over melt pool overturning and the melt pool surface composition evolves away from bulk pool composition and leads to  CaO-Al$_{2}$O$_{3}$ molten surfaces with geometric albedos of $\sim$0.5 \citep{rouan2011orbital}. Note that for  planets with substellar temperatures $<$~2400~K, melt-pool overturning circulation dominates over atmospheric transport of volatiles, and the melt pool surface composition is similar to the bulk pool composition. Rocky planets that may have CaO-Al$_{2}$O$_{3}$ molten surfaces include Kepler-10~b, Kepler-78~b and CoRoT-7~b (Figure \ref{RvsTemp_TOIs}).

One more controlling factor in the creation of evolved CaO-Al$_{2}$O$_{3}$ molten surfaces relates to the initial planetary bulk-silicate FeO concentration.  High initial FeO concentration promotes stratification, which hinders overturn, making evolved molten CaO-Al$_{2}$O$_{3}$ surfaces more likely \citep{kite2016atmosphere}.

\subsection{Reflection in Exoplanet Atmospheres} \label{reflectionatmospheres}

Atmospheres with reflective clouds may explain the high geometric albedos of some hot super Earths, provided the atmospheres have not been eroded.

A variety of atmospheric compositions have been proposed for hot super Earths, including CO and N$_{2}$, which are stable against dissociation at high temperatures \citep{angelo2017case}. Once all atmospheric volatiles have been removed, the silicate surface composition of the planet dictates the composition of the thin-silicate atmosphere that forms. High opaque cloud layers are predicted, primarily consisting of alkali metals and silicon oxides (e.g.  \citet{schaefer2009chemistry, miguel2011compositions, ito2015theoretical, kite2016atmosphere}).

Clouds consisting of alkalis and silicon oxide particles in the atmospheres of hot super Earths may result in high geometric albedos if the particles are smaller than the wavelength of the incident light. For spherical particles, scattering properties are determined mainly by the size of the particles and, to a lesser extent, their chemical composition. A particle's relative size (small/large)\ is dependent on the wavelength of light incident on the particle. As a particle becomes more forward scattering, the geometric albedo decreases. In general, large cloud particles (particle radius $\gg $ $\frac{\text{wavelength}}{\text{2}\pi }$) correspond to lower geometric albedos \citep{pierrehumbert2010principles}.

\subsection{Experimental Measurements of Specular and Diffuse Reflection From Molten Lava} \label{experimentslava}

In order to confirm the low albedo result for lava and quenched glass planetary surfaces, the reflection coefficient function for lava must be measured for multiple incidence and viewing angles. The bidirectional reflectance distribution function (BRDF)\ defines the surface reflection of light from a material. It is a function of the incident light direction and viewing direction, parametrized by the azimuth angle and zenith angle of both directions in a hemisphere \citep{hapke1981bidirectional}. The functional form of the BRDF of a material captures the specular and diffuse reflection components of the material.

BRDF measurements of our quenched glass samples were not deemed necessary because reflection along the specular reflection direction gives us an upper limit on the reflectance from the quenched glass samples at secondary eclipse, which is the aim of this study. This allowed for a simpler experimental setup, more repeatable experiments, and required a significantly smaller number of individual measurements, also making data representation more concise.

Measuring the albedo of molten lava requires several engineering challenges to be overcome. These challenges include: keeping lava molten and preventing a quenched glass crust from forming; efficiently obtaining several reflectance measurements of the molten lava at different incidence and viewing angles; and keeping the equipment at a safe distance from the lava, which complicates light source and spectrometer fiber alignment.  

Direct measurements of specular and diffuse reflection from molten lava would be useful to determine how appropriate values for molten silicates from non-crystalline solids literature are as reflectance proxies for lava.

\subsection{Future Prospects For The Discovery and Characterization of Lava Worlds} \label{detection}

There are a number of theories about the atmospheres and surfaces of lava-ocean planets. Characterizing the surfaces and atmospheres of hot super Earths in order to constrain sources of reflected light may help constrain these theories. The characterization of reflected light, however, requires bright host stars for high SNR data that has only recently become available.

Most lava-ocean planet candidates have been detected by \textit{Kepler}, making follow-up observations of these planets difficult since \textit{Kepler} planets orbit very faint stars ($<$ 12th magnitude). We now have the opportunity with \textit{TESS} to find more lava-ocean planet candidates amenable to follow-up observations, since \textit{TESS} surveys the nearest and brightest stars ($>$ 12th magnitude) \citep{ricker2014transiting}. \textit{TESS} is expected to find 556~$\pm $~31 planets with radii $<$ 2~R$_{\oplus }$ during its primary mission, some of which will be lava-ocean planet candidates \citep{sullivan2015transiting}.

\begin{figure*}[htb!]
\plotone{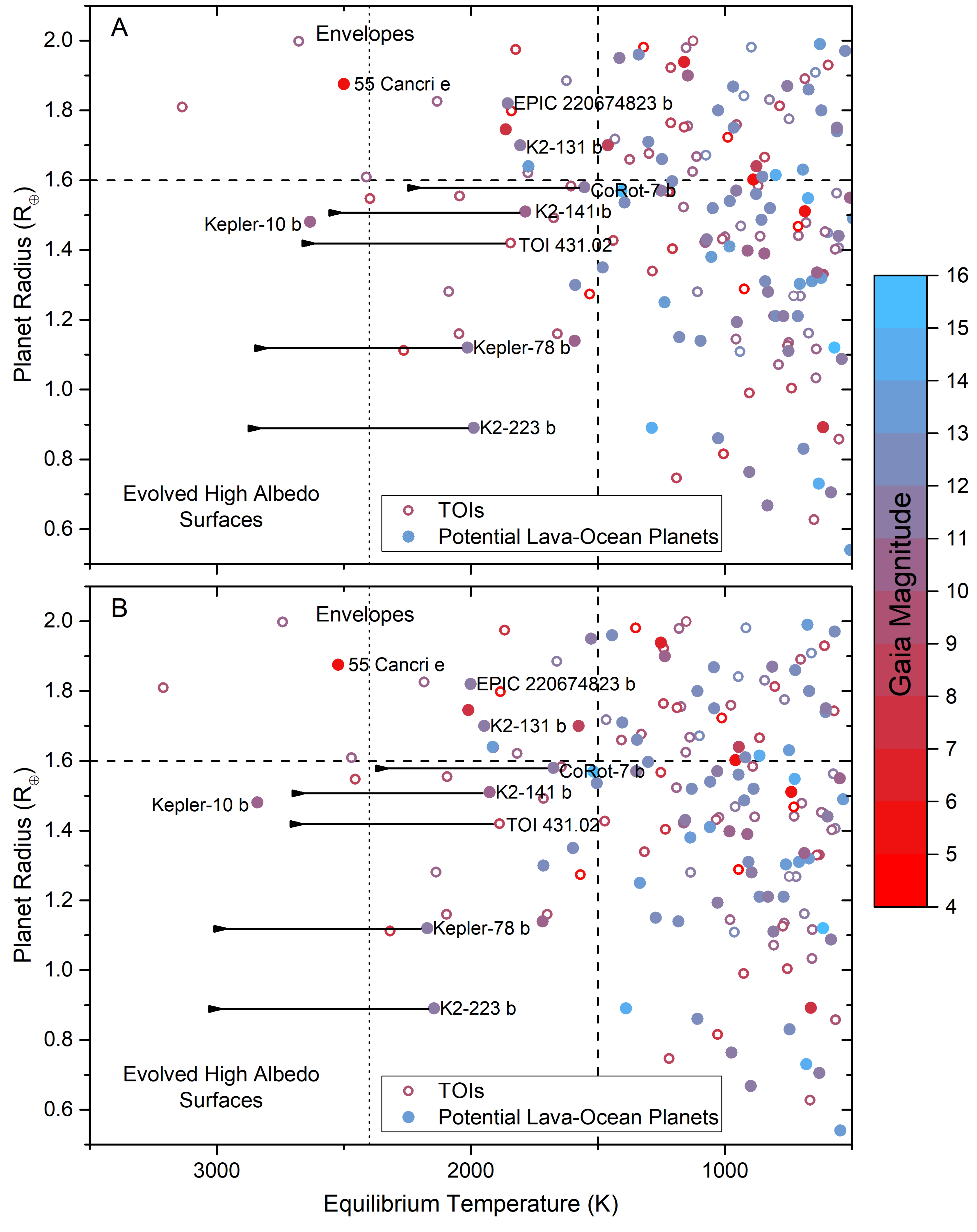}
\caption{Planet radius vs. equilibrium temperature of lava-ocean planet candidates. Lava-ocean planet candidates from the \textit{TESS} TOI list are denoted by open symbols. Planets are colored by GAIA magnitude of host star. (A) Equilibrium temperatures calculated assuming a basalt quenched glass albedo ($A_{g}$ = 0.09). (B) Equilibrium temperatures calculated assuming a feldspar quenched glass albedo ($A_{g}$ = 0.02). There is a 2\% difference between the calculated equilibrium temperatures using either basalt or feldspar quenched glass albedos. Planets with R$_{p}$ $>$ 1.6~R$_{\oplus }$ (horizontal dashed line) are likely to have volatile envelopes (\citet{rogers2015most, fulton2017california}). Planets with T$_{eq}$~$>$~2400~K (vertical dotted line) may have evolved high albedo surfaces \citep{kite2016atmosphere}. Planets with R$_{p}$ $<$ 1.6~R$_{\oplus }$ and T$_{eq}$~$<$~2400~K are expected to have bulk-silicate surfaces. For selected planets, the line attached to the data point represents the range of equilibrium temperatures for heat redistribution factor values ranging from $f=\frac{1}{4}$ (uniform heat redistribution) to $f=\frac{2}{3}$ (no heat redistribution/instant re-radiation). The most promising lava worlds will be those with T$_{eq}$~$>$~1500~K (vertical dashed line), encompassing a range of potential atmospheres and surface compositions.  \label{RvsTemp_TOIs}}
\end{figure*}

We identify potential lava-ocean planet candidates from the \textit{TESS}\ Objects of Interest (TOI) list\footnote{https://exoplanetarchive.ipac.caltech.edu}, as candidates for follow-up observations (Figure \ref{RvsTemp_TOIs}). 

In order to sustain lava at the substellar point, lava-ocean planet candidates must have substellar temperatures $>$~850~K. The equilibrium temperature of a planet, $T_{eq}$, is:

\begin{equation}
T_{eq}=T_{star}(1-A_{B})^{\frac{1}{4}}f^{\frac{1}{4}}\sqrt{\frac{R_{star}}{a}}\text{ ,}
\label{eqn15}
\end{equation}

\noindent where $T_{star}$ is the effective temperature of the star,$\ A$ is the Bond albedo, $f$ is the heat redistribution factor, $R_{star}$ is the stellar radius, and $a$ is the semi-major axis/orbital distance \citep{burrows2014spectra, de2015planetary}. 

In order to estimate the surface/substellar temperatures, $T_{sub}$, of the TOIs, we use equation (\ref{eqn15}), and assume a basalt/feldspar quenched glass albedo and uniform heat redistribution ($f=\frac{1}{4}$). Uniform heat redistribution is a conservative estimate to avoid excluding potential lava planet candidates based on atmospheric assumptions. We calculate a range of equilibrium temperatures for a few lava planet candidates by varying the heat distribution factor from $f=\frac{1}{4}$ (uniform heat redistribution) to $f=\frac{2}{3}$ (no heat redistribution/instant re-radiation) (Figure \ref{RvsTemp_TOIs}) \citep{spiegel2010atmosphere}.

$T_{sub}$ is calculated as \citep{mansfield2019identifying}:

\begin{equation}
T_{sub}=T_{eq}\left( \frac{1}{4}\right)^{-0.25}.
\label{eqn16}
\end{equation}

In order to characterize hot super Earths, the degeneracy between reflected light and thermal emission when interpreting the secondary eclipse depth of a planet in a single observation bandpass (Figure \ref{fig:AlbedoDegeneracy}) must be broken. This can be achieved by obtaining secondary eclipse depth measurements in two different, preferably overlapping, bandpasses \citep{placek2016combining}.

The James Webb Space Telescope (JWST), to be launched in 2021, has several spectroscopic instruments that can measure high resolution phase variations \citep{gardner2006james}. In particular, NIRCam has a wavelength range of 0.6$-$5~$\mu $m, which overlaps with the TESS band (600$-$1000~nm). NIRCam observations will contain a substantial thermal component but due to the overlap between the \textit{TESS} and NIRCam bandpasses, a change in albedo between observations will help to break the degeneracy between reflected light and thermal emission \citep{samuel2014constraining, placek2016combining}.

\section{Summary and Conclusion} \label{sec:conclusion}

There are a few hot super Earths that have observationally inferred high geometric albedo values ($>$ 0.4) in the Kepler band (420-900 nm). We considered molten lava and quenched glasses on the surfaces of hot super Earths as sources of reflected light that may contribute to the high albedo values. 

We experimentally measured reflectance from rough and smooth textured basalt and feldspar quenched glasses along the specular reflection direction. To supplement our measurements, we used specular reflection literature values of molten silicates as a proxy for specular reflectance values for lava. We integrated the reflectance values in a planet hemisphere model.

We found that reflection from rough and smooth quenched glasses and lava results in low planetary geometric albedos ($\lesssim$~0.1). Reflection from quenched glasses has the greatest contribution to reflectance at small incidence angles ($<$~20$^{\circ }$) and is negligible when considered over the entire planet. We conclude that lava worlds with solid (quenched glass) or liquid (lava) surfaces have low albedos, and the high geometric albedos of hot super Earths are likely explained by atmospheres with reflective clouds or, for a narrow range of parameter space, possibly Ca/Al oxide melt surfaces. 

The future of lava-ocean planet discovery and characterization lies in \textit{TESS}\ data and follow-up observations with instruments like \textit{JWST}.

\acknowledgments
We thank M. J. Tarkanian at the MIT Foundry, located in the Merton C. Flemings Materials Processing Laboratory, for help with producing the feldspar quenched glass. We thank the Syracuse University Lava Project for producing the basalt quenched glass. We thank N. Inamdar and R. Wei for help with precursor experiments and measurements. 
Z.E. thanks B. L. Ehlmann, C. R. Martin, and P. Niraula for their helpful discussions. Z.E. acknowledges funding for this research from the TESS mission and the Massachusetts Institute of Technology. 
This paper includes data collected by the TESS mission, which are publicly available from the Mikulski Archive for Space Telescopes (MAST). Funding for the TESS mission is provided by National Aeronautics and Space Administration’s (NASA) Science Mission directorate. This research has made use of the TESS Exoplanet Follow-up Observation Program website, which is operated by the California Institute of Technology, under contract with NASA under the Exoplanet Exploration Program. This research has made use of the NASA Exoplanet Archive, which is operated by the California Institute of Technology, under contract with NASA under the Exoplanet Exploration Program.

\facilities{TESS, MAST, Exoplanet Archive}
\software{Python, numpy, scipy, matplotlib, Origin}

\appendix
\renewcommand{\thefigure}{A\arabic{figure}}
\setcounter{figure}{0}

\section*{Reflectance Spectra of Quenched Glasses}

Spectrally-resolved reflectance measurements of our quenched glasses were converted into an average reflectance value for the visible wavelength range (Figures \ref{fig6}$-$\ref{fig10}).

The reflected counts data from the reference standard and quenched glass samples were binned in 10~nm intervals in order to reduce the noise in the data, after wavelength calibration. The reflected counts from the glass samples were divided  by  the  reflected  counts  from  the  reference standard at the corresponding incidence angle, and the reflectance value was calculated for  each incidence  angle. Finally, the spectrally-resolved reflectance measurements were averaged across the visible wavelength range (400$-$700~nm) corresponding to the wavelength range of the white LED illumination source, and to compare  to inferred albedo values from exoplanet observations. 

\begin{figure*}[htb!]\centering
\vspace*{-15mm}
\includegraphics[width = 4.89in, height = 9.7in]{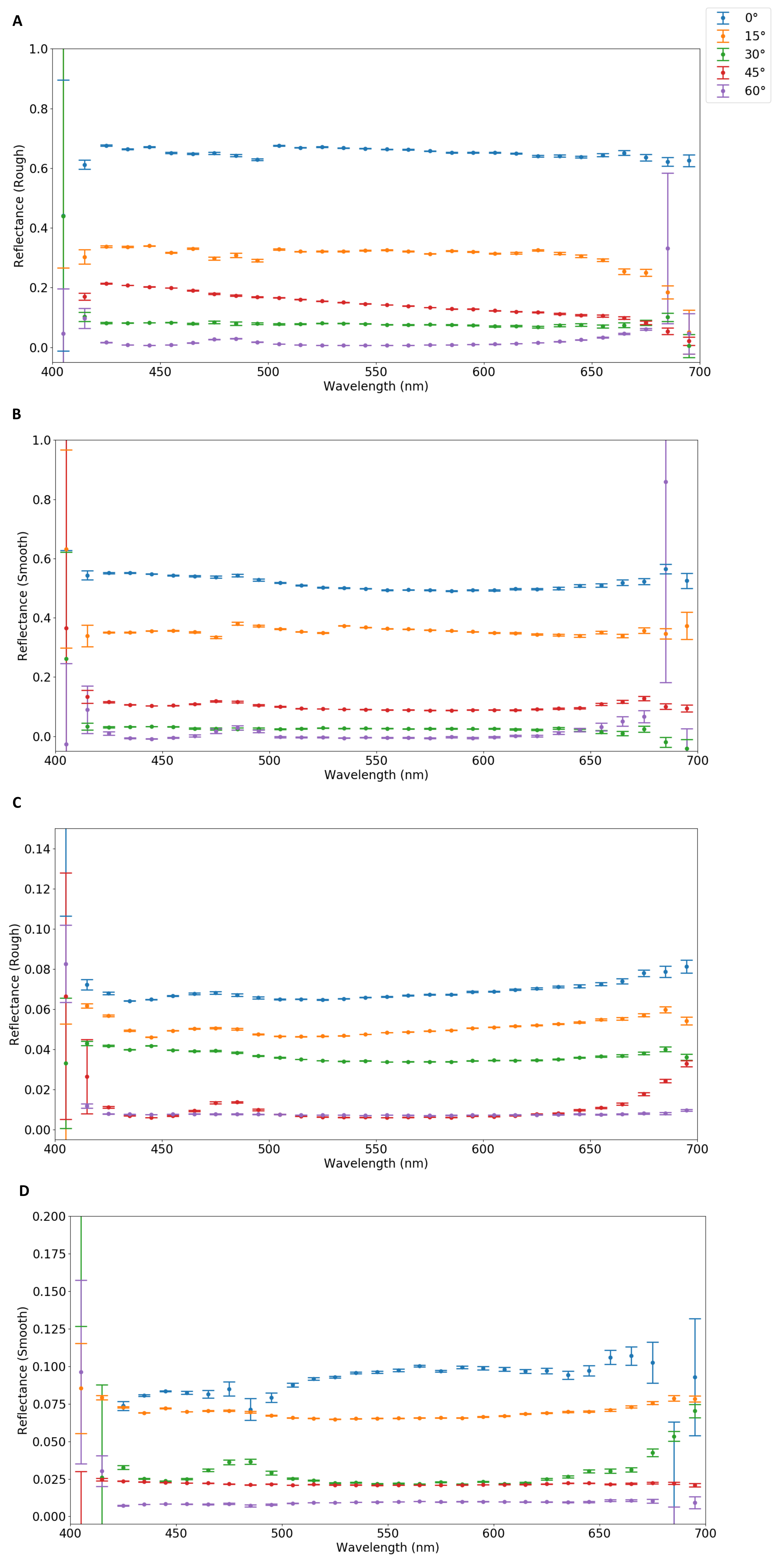}
\caption{Reflectance spectra of quenched glasses across the visible wavelength range (400$-$700~nm). Reflectance vs. wavelength values with 1$\protect\sigma $ errorbars plotted for incidence angles 0${{}^\circ}$ (blue), 15${{}^\circ}$ (orange), 30${{}^\circ}$ (green), 45${{}^\circ}$ (red) and 60${{}^\circ}$ (purple). (A) Rough basalt quenched glass. (B) Smooth basalt quenched glass. (C) Rough feldspar quenched glass. (D) Smooth feldspar quenched glass.}\label{ReflWavApx}
\end{figure*}

We find that reflectance does not vary significantly as a function of wavelength across the visible wavelength range (Figure \ref{ReflWavApx}). Some reflectance measurements exhibit increased noise but do not change significantly with wavelength. The shape of the reflectance curves does not change appreciably with incidence angle. The amplitude of the reflectance curve decreases with increasing incidence angle, as shown for the integrated wavelength values in Figures \ref{fig6}$-$\ref{fig10}. 

\begin{figure*}[htb!]\centering
\plotone{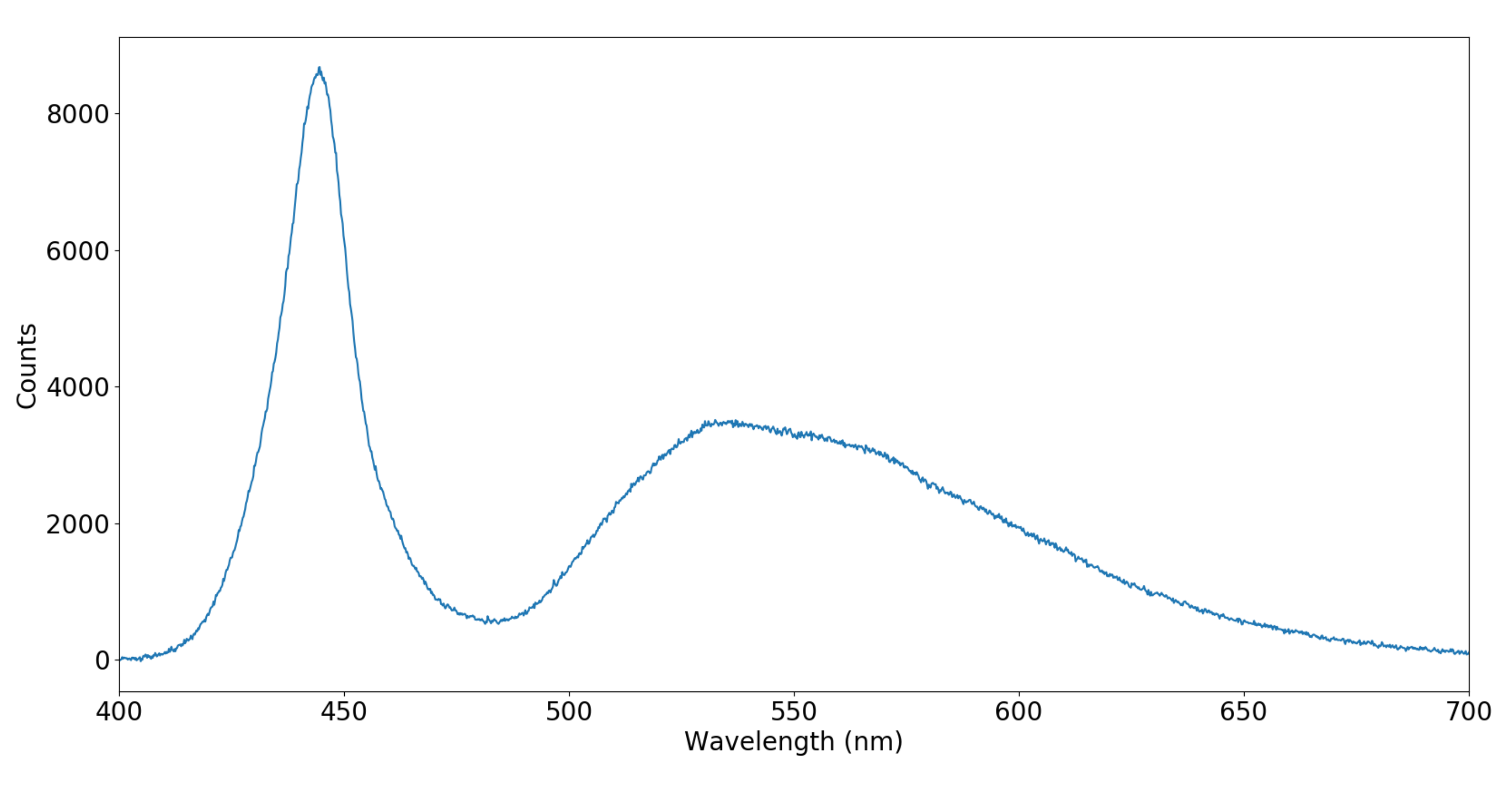}
\caption{White LED light source spectrum. The measured signal is close to/at the noise limit of the spectrometer at the beginning (400$-$410~nm) and end (680$-$700~nm) of the visible wavelength range. }\label{LEDspec}
\vspace*{5mm}
\end{figure*}

The increase in noise for reflectance values at the beginning and end of the visible wavelength range is systematic noise attributed to the white LED light source used to illuminate the quenched glass samples and reference standard. The light source does not emit enough light between 400$-$410~nm and 680$-$700~nm, leading to a measured signal close to/at the noise limit of the spectrometer (Figure \ref{LEDspec}). Since the measured signals are close to the noise limit at the beginning and end of the wavelength range, small fluctuations cause large changes in calculated reflectance values. The excess noise is present in both the individual spectra of the quenched glass samples and reference standard, implying that it is not a real effect unique to the glass samples. 

\FloatBarrier
\bibliography{lava_albedo}{}
\bibliographystyle{aasjournal}



\end{document}